\documentclass{emulateapj}
\shorttitle{The $M_{\rm bh}$--$L_{\rm spheroid}$ relation}
\shortauthors{Graham \& Scott}

\begin{document}

\title{The (black hole mass)--(host spheroid luminosity) relation at high and
  low masses, the quadratic growth of black holes, and intermediate-mass
  black hole candidates}

\author{Alister W.\ Graham\altaffilmark{1} and Nicholas Scott}
\affil{Centre for Astrophysics and Supercomputing, Swinburne University
of Technology, Hawthorn, Victoria 3122, Australia.}
\altaffiltext{1}{Corresponding Author: AGraham@swin.edu.au}

\begin{abstract} 

From a sample of 72 galaxies with reliable supermassive black hole masses
$M_{\rm bh}$, we derive the $M_{\rm bh}$--(host spheroid luminosity, $L$)
relation for {\it i)} the subsample of 24 core-S\'ersic galaxies with
partially depleted cores, and {\it ii)} the remaining subsample of 48 S\'ersic
galaxies.
Using $K_s$-band {\it Two Micron All Sky Survey} data, we find the near-linear
relation $M_{\rm bh} \propto L_{\rm K_s}^{1.10\pm0.20}$ for the core-S\'ersic
spheroids thought to be built in additive dry merger events, while $M_{\rm bh}
\propto L_{\rm K_s}^{2.73\pm0.55}$ for the S\'ersic spheroids built from
gas-rich processes.
After converting literature $B$-band disk {\it galaxy} magnitudes into
inclination- and dust-corrected {\it bulge} magnitudes, via a useful new
equation presented herein, we obtain a similar result.
Unlike with the $M_{\rm bh}$--(velocity dispersion) diagram, which is also
updated here using the same galaxy sample, it remains unknown whether barred
and non-barred S\'ersic galaxies are offset from each other in the $M_{\rm
bh}$--$L$ diagram.

While black hole feedback has typically been invoked to explain 
what was previously thought to be a nearly constant $M_{\rm
bh}/M_{\rm Spheroid}$ mass ratio of $\sim$0.2\%, we advocate 
that the near-linear $M_{\rm bh}$--$L$ and $M_{\rm bh}$--$M_{\rm Spheroid}$
relations observed at high masses may have instead largely arisen from the 
additive dry merging of galaxies.  We argue that 
feedback results in a dramatically different scaling relation, such that 
black hole mass scales 
roughly quadratically with the spheroid mass in S\'ersic galaxies. 
%
We therefore 
introduce a revised cold-gas 'quasar' mode feeding equation for 
semi-analytical models to reflect what we dub the {\it quadratic growth} of
black holes in S\'ersic galaxies built amidst gas-rich processes. 
Finally, we use our new S\'ersic $M_{\rm bh}$--$L$ equations to predict the
masses of candidate {\it intermediate mass} black holes in almost 50 low
luminosity spheroids containing active galactic nuclei, finding 
many masses between that of stellar mass black holes and supermassive black holes.

\end{abstract}

\keywords{ 
black hole physics --- 
galaxies: evolution ---
galaxies: nuclei
}
  
\section{Introduction} 

The (black hole mass)--(host spheroid dynamical mass) $M_{\rm bh}$--$M_{\rm
Sph,dyn}$ relation (e.g.\ Marconi \& Hunt 2003; H\"aring \& Rix 2004) 
has recently been revised such that 
core-S\'ersic galaxies (whose spheroidal components are thought to be built in
simple additive dry merger events) define a near-linear relation while
S\'ersic galaxies (whose spheroidal components are likely built amidst
gas-rich processes) define a near-quadratic relation (Graham 2012a).
Graham (2012a,b) additionally predicted that the previously established
log-linear, i.e.\ single power-law, relation between black hole mass and host
spheroid luminosity, $L$, must also require a significant modification such
that $M_{\rm bh} \propto L^{1.0}$ for the spheroidal component of luminous
($M_B \lesssim -20.5$ mag) core-S\'ersic galaxies while $M_{\rm bh} \propto
L^{2.5}$ for the fainter S\'ersic spheroids --- if their dynamical
mass-to-light ratio scales with $L^{1/4}$.  If $(M/L)_{\rm dynamical} \propto
L^{1/3}$ then one expects $M_{\rm bh} \propto L^{8/3}$.  Here, for the first
time, we show that the $M_{\rm bh}$--$L$ relation is indeed bent, and we
provide new expressions to predict black hole masses in core-S\'ersic galaxies
and S\'ersic galaxies using the luminosity of their spheroidal
component.\footnote{Readers unfamiliar with the core-S\'ersic model (Graham et
al.\ 2003; Trujillo et al.\ 2004) or the S\'ersic (1963) model, for the
classification of galaxies, may like to refer to the review article by Graham
\& Driver (2005).}

From a literature sample of eight galaxies, Kormendy \& Richstone (1995, their
figure~14)
revealed an apparently linear correlation between black hole mass and the
brightness / mass of the host spheroid.  As the sample size grew, researchers
began to apply a single log-linear, i.e.\ power-law, relation to the data
(e.g.\ Richstone et al.\ 1998; Kormendy \& Gebhardt 2001; McClure \& Dunlop
2002).  However 
%
%
the samples remained dominated by fairly luminous spheroids and the
applicability of the new $M_{\rm bh}$--$L$ relations (e.g.\ McClure \& Dunlop
2004; Marconi \& Hunt 2003; Graham 2007a) to low-luminosity systems was not
securely established. 
Indeed, Graham (2007a) noted that the $M_{\rm bh}$--$L$ and $M_{\rm
bh}$--$\sigma$ relations (Ferrarese \& Merritt 2000; Gebhardt et al.\ 2000)
cannot both be described by a power-law because the relationship between
luminosity and velocity dispersion, $\sigma$, for elliptical galaxies is not
described by a single power-law.

As reviewed in Graham (2012c), the elliptical galaxy $L$--$\sigma$ relation
has long been known to be `broken', such that $L \propto \sigma^5$ at the
luminous-end ($M_B \lesssim -20.5$ mag: Schechter 1980; Malumuth \& Kirshner
1981; Von Der Linden et al.\ 2007; Liu et al.\ 2008) and possibly even as
steep as $L \propto \sigma^6$ (Lauer et al.\ 2007),
while $L \propto \sigma^2$ at intermediate and faint luminosities (e.g.\
Davies et al.\ 1983; Held et al.\ 1992; de Rijcke et al.\ 2005; Matkovi\'c \&
Guzm\'an 2005; 
Kourkchi et al.\ 2012).  Galaxy samples containing both bright
and intermediate luminosity elliptical galaxies will therefore naturally yield $L$--$\sigma$ 
relationships with exponents around 3 to 4 (e.g., Faber \& Jackson 1976; Tonry 1981; de
Vaucouleurs \& Olson 1982; Desroches et al.\ 2007).
It should be noted that this change in slope of the $L$--$\sigma$
relation at $M_B \approx -20.5$ mag 
(S\'ersic $n \approx$ 3--4)  
is not related to (i) the onset of the coexistence of classical bulges and
pseudobulges at $n_{\rm bulge} \lesssim 2$, nor (ii) 
the alleged 
%
divide at $M_B = -18$ mag ($n_{\rm galaxy} \lesssim$ 1--2) between dwarf and
ordinary elliptical galaxies (Kormendy 1985; Kormendy et al.\ 2009).  Instead,
the change in slope at $M_B \approx -20.5$ mag coincides with the division
between core-S\'ersic galaxies (which have a central deficit of light relative
to the inward extrapolation of their outer S\'ersic light profile) and
S\'ersic galaxies\footnote{S\'ersic galaxies were previously referred to as
`power-law' galaxies before it was realized that their spheroidal component's
inner light profiles are better described by the curved S\'ersic model than by
a power-law (e.g.\ Trujillo et al.\ 2004).}  which do not (Graham \& Guzm\'an
2003; Graham et al.\ 2003; Trujillo et al.\ 2004; Matkovi\'c \& Guzm\'an
2005).  If (non-barred) S\'ersic and core-S\'ersic galaxies follow the same
$M_{\rm bh} \propto \sigma^5$ relation (see section~\ref{SecMs}), they
therefore obviously cannot follow the same $M_{\rm bh}$--$L$ relation because
they do not follow the same $L$--$\sigma$ relation.  Simple consistency
arguments dictate that the core-S\'ersic galaxies should follow the relation
$M_{\rm bh} \propto L^{1.0}$ while the (non-barred) S\'ersic galaxies should follow the
relation $M_{\rm bh} \propto L^{2.5}$ (or $M_{\rm bh} \propto L^3$ if $M_{\rm
bh} \propto \sigma^6$).


While the core-S\'ersic model has provided a new means to identify and quantify 
the stellar deficits, i.e.\ the flattened cores, which have long been observed
in giant galaxies (e.g.\ King 1978, and references therein), the rationale for
a physical divide between core-S\'ersic galaxies and S\'ersic galaxies 
also existed long ago. 
Early-type galaxies brighter than $M_B = -20.5$ mag tend to be anisotropic,
pressure supported elliptical galaxies with boxy isophotes, while the less
luminous early-type galaxies tend to have disky isophotes and often contain a
rotating disk of stars (e.g.\ Carter 1978, 1987; Davies et al.\ 1983; Bender
et al.\ 1988; Peletier et al.\ 1990; Jaffe et al.\ 1994; Faber et al.\ 1997; 
Emsellem et al.\ 2011). 
In addition, and as noted already, the $L$--$\sigma$ relation changes slope at
$M_B \approx -20.5$ mag (Davies et al.\ 1983; Matkovi\'c \& Guzm\'an 2005):
the core-S\'ersic galaxies ($M_B \lesssim -20.5$ mag) define the relation $L
\propto \sigma^5$ while the S\'ersic galaxies ($M_B \gtrsim -20.5$ mag) follow
the relation $L \propto \sigma^2$, with no discontinuity at $M_B = -18$ mag.
Similarly, the elliptical galaxy $L$--$\mu_0$ relation, 
where $\mu_0$ is the central surface brightness, also breaks at $M_B \approx
-20.5$ mag and remains linear at $M_B = -18$ mag (e.g.\ Graham \&
Guzm\'an 2003)\footnote{As shown, and explained, in Graham 2012c, it is only
from the use of `effective' radii and `effective' surface brightnesses that apparent
breaks are seen at $M_B = -18$ mag in what are actually continuous, curved relations.}.
While gas-rich processes and mergers are thought to build the S\'ersic
galaxies, relatively gas-free galaxy mergers are thought to build the
core-S\'ersic galaxies. The coalescence of massive black holes --- from the
pre-merged, gas-poor galaxies --- scour out the core of the newly merged
`core-S\'ersic'  galaxy 
(Begelman, Blandford, \& Rees 1980; Ebisuzaki, Makino, \& Okumura 1991;
Merritt, Mikkola, \& Szell 2007). 

The above expectation for the `bent' $M_{\rm bh}$--$L$ relation 
impacts dramatically upon our predictions for black hole
masses at the centers of other galaxies, has implications for the competing
formation scenarios proposed to explain the coevolution of galactic spheroids
and their supermassive black holes, and has ramifications for semi-analytic
computer codes which try to mimic this coevolution.  Moreover, as noted in
Graham (2012a), the realization of a broken $M_{\rm bh}$--$L$ relation will
influence, among other things: evolutionary studies of the (black hole)-(host
spheroid) connection over different cosmic epochs (e.g.\ 
Cisternas et al.\ 2011; 
Li, Ho \& Wang 2011; 
Hiner et al.\ 2012; 
Zhang, Lu \& Yu 2012); predictions for space-based gravitational wave
detections which use the $M_{\rm bh}$--$L$ relation to estimate black hole
masses (e.g.\ Mapelli et al.\ 2012, and references therein); and estimates of
the black hole mass function and mass density based on spheroid luminosity
functions (e.g.\ Vika et al.\ 2009; Li, Wang, \& Ho 2012).

Recent studies of the $M_{\rm bh}$--$L$ relation have continued to be largely
dominated by luminous spheroids with $M_{\rm bh} \gtrsim 10^7$--$10^8
M_{\odot}$, and have continued to fit a straight line to the data (e.g.\ Sani
et al.\ 2011; Vika et al.\ 2011; McConnell et al.\ 2011a; Beifiori et al.\ 2012).  However, an
inspection of their $M_{\rm bh}$--$L$ diagrams reveals the
onset of a steepening relation in their lower luminosity spheroids which have $M_{\rm
  bh} \lesssim 10^7$--$10^8 M_{\odot}$, as also seen in the 
diagrams from Graham (2007a) and G\"ultekin et al.\ (2009b). 
In addition, a number of models are actually generating `bent' 
  $M_{\rm bh}$--$M_{\rm Spheroid}$ relations which show a steepening at low
  masses (e.g.\ Cirasuolo et al.\ 2005, their figure~5;
  Khandai et al.\ 2012, their figure~7; 
  Bonoli, Mayer \& Callegari 2012, their figure~7). 
Here we build on
this (tentative observational) evidence of a steepening relation at low 
masses by using a sample of 72 galaxies with directly measured black hole
masses that are tabulated in Section~2 and span $10^6$--$10^{10} M_{\odot}$.
For the first time we both identify {\it and} quantify the bend in the $M_{\rm
bh}$--$L$ relation (Section~3).

Rather than fitting a single quadratic relation to the ($M_{\rm bh}, L$)
distribution of points --- which would be in line with the use of a 
log-quadratic relation in the $M_{\rm bh}$--$n$ diagram by Graham \& Driver
(2007a) --- we embrace the large body of data which shows that S\'ersic and
core-S\'ersic galaxies have distinguishing characteristics suggestive of a
different formation history (e.g.\ Davies et al.\ 1983; Faber et al.\ 1997;
Graham \& Guzm\'an 2003; Gavazzi et al.\ 2005; Ferrarese et al.\ 2006) and
therefore fit two $M_{\rm bh}$--$L$ relations, one for each type of galaxy. 

 
In Section~4 we provide a discussion of
some of the more salient points arising from the new $M_{\rm
bh}$--$L$ relation for S\'ersic and core-S\'ersic galaxies. 
In particular, Section~4.1 provides updated black hole mass estimates for nearly 50
low-luminosity spheroids, and reveals that they occupy the holy grail mass
range for intermediate mass black holes, ranging from the 
stellar mass black hole upper limit of 
$\sim$$10^2 M_{\odot}$ to the lower-bound of $10^5$--$10^6
M_{\odot}$ for supermassive black holes. 
In Section~4.2 we introduce a significantly modified 
expression for use in semi-analytical models which try to mimic the
coevolution of supermassive black holes and galaxies.  We provide a new 
{\it quadratic} cold-gas `quasar' mode feeding equation to match the quadratic
(black hole)-to-(host spheroid) growth observed in S\'ersic galaxies built
from gas-rich processes. 
Finally, Section~4.3 discusses the observation that barred and non-barred disk
galaxies currently appear to occupy the same region of the $M_{\rm bh}$--$L$
diagram.

\section{Data} \label{Sec_Data}

We have continued to build on past catalogs of reliable supermassive black
hole masses obtained from direct maser, stellar or gas kinematic measurements 
(e.g.\ Ferrarese \& Ford (2005); Hu (2008); Graham (2008b); Graham et
al.\ 2011) by adding 16 galaxies from the literature.  This gives us an 
initial sample of 80 galaxies with directly measured black hole masses.
Since the catalog in Graham et al.\ (2011) was prepared, 
we note that several updates over the last two years have resulted 
in the doubling of black hole masses at the high-mass end of the $M_{\rm
bh}$--$\sigma$ diagram (e.g.\ van den Bosch \& de Zeeuw 2010; Walsh et al.\
2010, 2012) and two ten-billion solar mass black holes have been reported 
(e.g.\ McConnell et al.\ 2012). 

\subsection{Galaxy exclusions}

Separate from the above sample of 80 galaxies, 
we have maintained our exclusion of NGC~7457, because its black hole mass was derived
assuming that this galaxy's excess nuclear light was due to an AGN rather than
a massive nuclear disk and dense star cluster (Balcells et al.\ 2007; Graham \&
Spitler 2009), and the exclusion of the Sc galaxy NGC~2748 due to potential dust
complications with its estimated black hole mass (Atkinson et al.\ 2005; Hu
2008). 

For the following reasons, 
our sample of 80 galaxies is reduced here by five. 
The best-fitting parameter set in the mass modeling of NGC~5252 by
Capetti et al.\ (2005) had resulted in a reduced $\chi^2$ value of 16.5,
indicative of a poor fit to the data and resulting in its exclusion by
G\"ultekin et al.\ (2009b).  We too now exclude this galaxy, which would
otherwise appear to have a black hole mass that is an order of magnitude too
high in the $M_{\rm bh}$--$\sigma$ diagram.
The inclined, starburst spiral galaxy Circinus is also excluded 
due to its complex kinematics (e.g.\ For, Koribalski \& Jarrett 2012, and
references therein) and location on the other side of the Galactic plane.  With a
Galactic latitude of less than 5 degrees, the suggested $B$-band extinction
correction for Circinus is 5 mag or more (Schlafly \& Finkbeiner
2011) and is considered to be unreliable.
The above two galaxies can not be used reliably in either the $M_{\rm
bh}$--$L$ diagram or the $M_{\rm bh}$--$\sigma$ diagram, and they 
are therefore not listed in our Table~\ref{Tab1}. 

Following G\"ultekin et al.\ (2009b), we exclude the barred galaxy
NGC~3079. 
Although NGC~3079 has a well determined black hole mass (Kondratko et al.\
2005), the observed stellar velocity dispersion drops rapidly from a central
value of $\sim$150 km s$^{-1}$ to 60 km s$^{-1}$ or less within just 
1 kpc (Shaw, Wilkinson, \& Carter 1993).  The dynamics in this Sc
galaxy's boxy peanut-shaped bulge are known to be affected by bar streaming
motions (Merrifield \& Kuijken (1999; Veilleux, Bland-Hawthorn, \& Cecil 1999)
which were not modeled by Shaw et al.\ (1993).  Such additional motions are
a general concern in barred galaxies, as detailed in Graham et al.\ (2011),
and it is not clear what velocity dispersion should be used for the bulge
component of this galaxy\footnote{While this galaxy could be used in the
$M_{\rm bh}$--$L_{\rm spheroid}$ diagram, we do not so as to be consistent
with our galaxy sample in the final $M_{\rm bh}$--$\sigma$ diagram.}.
The bulge luminosity of the Milky Way is uncertain due to dust in
our Galaxy's disc, and is therefore not included in our $M_{\rm bh}$--$L_{\rm
  spheroid}$ diagram. 
%
Finally, we also exclude M32 which belongs to the rare `compact elliptical'
galaxy class and is thus not representative of the majority of galaxies.  
Such galaxies are thought to be heavily stripped of an unknown fraction
of their stars (e.g.\ Bekki et al.\ 2001), and as such this galaxy may bias our analysis. 
We do however note that M32, and the Milky Way, have been included in a preliminary
$M_{\rm bh}$--$\sigma$ diagram (Figure~\ref{Fig1}).  We therefore mention here that 
we have used a host bulge velocity dispersion of 55 km s$^{-1}$ 
from M32's bulge stars outside of M32's core region; this value 
is lower than the commonly used central aperture 
values of 72--75 km s$^{-1}$ which are biased high by the dynamics around the
central black hole (Igor Chilingarian, priv.\ comm.).
This leaves us, thus far, with a sample of ($80-5 = $) 75 galaxies.  
For reasons discussed at the start of Section~\ref{SecAaR}, 
an additional three galaxies (NGC~1316, NGC~3842 and NGC~4889) 
are excluded from some of the analyses.

\subsection{Galaxy distances and magnitudes}

For the bulk of the sample, we have used the distance moduli from Tonry et al.\ (2001) 
after first decreasing these values by 0.06 mag and thereby reducing both the
galaxy distances and the black hole masses by $\sim$3\%.  This small adjustment
arises from Blakeslee et al.'s (2002, their Section~4.6) recalibration of the
surface brightness fluctuation method based on the final Cepheid distances
given by Freedman et al.\ (2001, with the metallicity correction).
In addition to listing each galaxy's distance and velocity dispersion, 
primarily taken from HyperLeda\footnote{http://leda.univ-lyon1.fr} (Paturel et al.\ 2003), 
Table~\ref{Tab1} presents their observed (i.e.\ uncorrected) 
$B$-band magnitude, $B_T$, from the {\it Third Reference Catalogue of Bright
Galaxies} (de Vaucouleurs et al.\ 1991, hereafter RC3) and their total
$K_s$-band magnitude as provided by the {\it Two Micron All Sky Survey} 
(2MASS)\footnote{www.ipac.caltech.edu/2mass} (Jarrett et al.\ 2000).  
Due to the bright star near the center of NGC~2974, we used the $K$-band
magnitude from Cappellari et al.\ (2006) rather than the 2MASS value for this
galaxy.  

Both the observed $B$- and $K_s$-band 
magnitudes need to be corrected for Galactic extinction, which we have taken
from Schlegel et al.\ (1998), as provided by the 
NASA/IPAC Extragalactic Database 
(NED)\footnote{http://nedwww.ipac.caltech.edu} and included in Table~1 as $A_B$
and $A_K$ for ease of reference.  
We additionally provide K-corrections for the 2MASS $K_s$-band
magnitudes, derived 
from each galaxy's heliocentric redshift $z$ (taken from NED) and total $J-K$ color 
(taken from 2MASS) coupled with 
the prescription from Chilingarian, Melchior \& Zolotukhin
(2010) which is available online\footnote{http://kcor.sai.msu.ru}.
In passing, we note that for the low redshifts associated with our galaxy
sample, this K$_{K_s}$ correction is approximately $-2.1z$, $-2.2z$, and
$-2.5z$ for the elliptical, S0 and Sa, and later spiral galaxy types, respectively.
Without reliable $B-R$ or $B-I$ colors, the $B$-band K-corrections tabulated in Table~1
are such that K$_{B}$ equals $4z$, $3.5z$, $3z$, and $2.5z$ for the 
elliptical, S0, Sa, and later spiral galaxy types, respectively.
As can be seen in Table~\ref{Tab1}, these corrections are minor and they have had
no significant impact on the final results. 
Finally, in our analysis we corrected all galaxy magnitudes for cosmological 
redshift dimming.


\scriptsize{

\begin{deluxetable*}{llllllllccccccc}
\tablewidth{420pt}
\tablecaption{Black hole / Galaxy data. \label{Tab1}} 
\tablehead{ 
\colhead{Galaxy} & \colhead{Type} & \colhead{Dist} & \colhead{core} & \colhead{$\sigma$} & \colhead{$M_{\rm bh}$} & 
\colhead{$B_T$} & \colhead{$K_s$} & \colhead{A$_B$} & \colhead{A$_K$} & \colhead{K$_B$} & \colhead{K$_{K_s}$} & \colhead{R$_{25}$} 
& \colhead{M$_B$}  &  \colhead{M$_{K_s}$}  \\
\colhead{ } & \colhead{ } & \colhead{[Mpc]} & \colhead{ } & \colhead{km s$^{-1}$} & \colhead{[10$^8 M_{\odot}$]} & 
\colhead{[mag]} & \colhead{[mag]} & \colhead{[mag]} & \colhead{[mag]} & \colhead{[mag]} & \colhead{[mag]} & \colhead{ } 
&  \colhead{[mag]} & \colhead{[mag]}  \\
\colhead{(1)} & \colhead{(2)} & \colhead{(3)} & \colhead{(4)} & \colhead{(5)} & \colhead{(6)} & \colhead{(7)} & 
\colhead{(8)} & \colhead{(9)} & \colhead{(10)} & \colhead{(11)} & \colhead{(12)} & \colhead{(13)} & \colhead{(14)} & \colhead{(15)} 
}
\startdata 
\multicolumn{15}{c}{30 Ellipticals} \\
A1836, BCG& E1      & 158.0 [3a]& y?    & 309 [5a] &   39$^{+    4}_{-    5}$ [6a]  &  14.56      &  9.993  &  0.277 &  0.024 & 0.15 & $-$0.07 &   ... &  $-$21.72  &  $-$26.25 \\
A3565, BCG& E1      &  40.7 [3a]& y?    & 335 [5b] &   11$^{+    2}_{-    2}$ [6a]  &  11.61      &  7.502  &  0.265 &  0.023 & 0.05 & $-$0.03 &   ... &  $-$21.71  &  $-$25.65 \\
IC 1459   & E3      &  28.4     & y?    & 306      &   24$^{+   10}_{-   10}$ [6b]  &  11.61      &  7.502  &  0.265 &  0.023 & 0.02 & $-$0.01 &   ... &  $-$21.37  &  $-$25.50 \\
NGC  821  & E6      &  23.4     & n [4a]& 200      & 0.39$^{+ 0.26}_{- 0.09}$ [6c]  &  11.67      &  7.900  &  0.474 &  0.040 & 0.02 & $-$0.01 &   ... &  $-$20.65  &  $-$24.02 \\
NGC 1399  & E1      &  19.4     & y [4b]& 329      &  4.7$^{+  0.6}_{-  0.6}$ [6d]  &  10.55      &  6.306  &  0.056 &  0.005 & 0.02 & $-$0.01 &   ... &  $-$20.95  &  $-$25.17 \\
NGC 2974  & E4      &  20.9     & n [4c]& 227      &  1.7$^{+  0.2}_{-  0.2}$ [6e]  &  11.87      &8.00 [8a]&  0.235 &  0.020 & 0.03 & $-$0.01 &   ... &  $-$19.97  &  $-$23.66 \\ 
NGC 3377  & E5      &  10.9     & n [4d]& 139      & 0.77$^{+ 0.04}_{- 0.06}$ [6f]  &  11.24      &  7.441  &  0.147 &  0.013 & 0.01 & $-$0.01 &   ... &  $-$19.09  &  $-$22.78 \\    
NGC 3379  & E1      &  10.3     & y [4b]& 209      &  4.0$^{+  1.0}_{-  1.0}$ [6g]  &  10.24      &  6.270  &  0.105 &  0.009 & 0.01 & $-$0.01 &   ... &  $-$19.93  &  $-$23.83 \\    
NGC 3608  & E2      &  22.3     & y [4b]& 192      &  2.0$^{+  1.1}_{-  0.6}$ [6c]  &  11.70      &  8.096  &  0.090 &  0.008 & 0.02 & $-$0.01 &   ... &  $-$20.13  &  $-$23.68 \\    
NGC 3842  & E3      &  98.4 [6h]& y [4b]& 270 [6h] &   97$^{+   30}_{-   26}$ [6h]  &  12.78      &  9.082  &  0.093 &  0.008 & 0.08 & $-$0.04 &  ...  &  $-$22.28  &  $-$26.02 \\
NGC 4261  & E2      &  30.8     & y [4e]& 309      &  5.0$^{+  1.0}_{-  1.0}$ [6i]  &  11.41      &  7.263  &  0.078 &  0.007 & 0.03 & $-$0.02 &   ... &  $-$21.11  &  $-$25.24 \\    
NGC 4291  & E3      &  25.5     & y [4b]& 285      &  3.3$^{+  0.9}_{-  2.5}$ [6c]  &  12.43      &  8.417  &  0.158 &  0.013 & 0.02 & $-$0.01 &   ... &  $-$19.76  &  $-$23.66 \\    
NGC 4374  & E1      &  17.9     & y [4f]& 296 [6j] &  9.0$^{+  0.9}_{-  0.8}$ [6j]  &  10.09      &  6.222  &  0.174 &  0.015 & 0.01 & $-$0.01 &   ... &  $-$21.35  &  $-$25.08 \\    
NGC 4473  & E5      &  15.3     & n [4b]& 179      &  1.2$^{+  0.4}_{-  0.9}$ [6c]  &  11.16      &  7.157  &  0.123 &  0.010 & 0.03 & $-$0.02 &   ... &  $-$19.89  &  $-$23.83 \\    
NGC 4486  & E1      &  15.6     & y [4f]& 334      &   58$^{+  3.5}_{-  3.5}$ [6k]  &  09.59      &  5.812  &  0.096 &  0.008 & 0.02 & $-$0.01 &   ... &  $-$21.47  &  $-$25.19 \\ 
NGC 4486a & E2      &  17.0 [3b]& n [4f]& 110 [5c] & 0.13$^{+ 0.08}_{- 0.08}$ [6l]  &  13.20 [7a] &  9.012  &  0.102 &  0.009 & 0.00 & $-$0.00 &   ... &  $-$18.05  &  $-$22.15 \\
NGC 4552  & E0      &  14.9     & y [4b]& 252      &  4.7$^{+  0.5}_{-  0.5}$ [6e]  &  10.73      &  6.728  &  0.177 &  0.015 & 0.00 & $-$0.00 &   ... &  $-$20.31  &  $-$24.16 \\
NGC 4621  & E5      &  17.8     & n [4f]& 225      &  3.9$^{+  0.4}_{-  0.4}$ [6e]  &  10.57      &  6.746  &  0.143 &  0.012 & 0.01 & $-$0.00 &   ... &  $-$20.83  &  $-$24.52 \\
NGC 4649  & E2      &  16.4     & y [4b]& 335      &   47$^{+   10}_{-   10}$ [6m]  &   9.81      &  5.739  &  0.114 &  0.010 & 0.01 & $-$0.01 &   ... &  $-$21.38  &  $-$25.37 \\    
NGC 4697  & E6      &  11.4     & n [4g]& 171      &  1.8$^{+  0.2}_{-  0.1}$ [6c]  &  10.14      &  6.367  &  0.131 &  0.011 & 0.02 & $-$0.01 &   ... &  $-$20.28  &  $-$23.96 \\    
NGC 4889  & E4      & 103.2 [6h]& y [4g]& 347 [6h] &  210$^{+  160}_{-  160}$ [6h]  &  12.53      &  8.407  &  0.041 &  0.004 & 0.09 & $-$0.04 &  ...  &  $-$22.59  &  $-$26.80 \\
NGC 5077  & E3      &  41.2 [3a]& y [4h]& 255      &  7.4$^{+  4.7}_{-  3.0}$ [6n]  &  12.38      &  8.216  &  0.210 &  0.018 & 0.04 & $-$0.02 &   ... &  $-$20.91  &  $-$24.94 \\
NGC 5576  & E3      &  24.8     & n [4b]& 171      &  1.6$^{+  0.3}_{-  0.4}$ [6o]  &  11.85      &  7.827  &  0.136 &  0.012 & 0.02 & $-$0.01 &   ... &  $-$20.26  &  $-$24.19 \\
NGC 5813  & E1      &  31.3     & y [4b]& 239      &  6.8$^{+  0.7}_{-  0.7}$ [6e]  &  11.45      &  7.413  &  0.246 &  0.021 & 0.03 & $-$0.01 &   ... &  $-$21.28  &  $-$25.12 \\
NGC 5845  & E4      &  25.2     & n [4h]& 238      &  2.6$^{+  0.4}_{-  1.5}$ [6c]  &  13.50      &  9.112  &  0.230 &  0.020 & 0.02 & $-$0.01 &   ... &  $-$18.74  &  $-$22.95 \\    
NGC 5846  & E0      &  24.2     & y [4i]& 237      &   11$^{+    1}_{-    1}$ [6e]  &  11.05      &  6.949  &  0.237 &  0.020 & 0.02 & $-$0.01 &   ... &  $-$21.11  &  $-$25.02 \\ 
NGC 6086  & E3      & 138.0 [3a]& y [4j]& 318 [6p] &   37$^{+   18}_{-   11}$ [6p]  &  13.79      &  9.973  &  0.162 &  0.014 & 0.13 & $-$0.05 &  ...  &  $-$22.08  &  $-$25.93 \\
NGC 6251  & E2      & 104.6 [3a]&y? [6q]& 311      &  5.9$^{+  2.0}_{-  2.0}$ [6q]  &  13.64 [7b] &  9.026  &  0.377 &  0.032 & 0.10 & $-$0.04 &   ... &  $-$21.84  &  $-$26.25 \\
NGC 7052  & E4      &  66.4 [3a]& y [4k]& 277      &  3.7$^{+  2.6}_{-  1.5}$ [6r]  &  13.40 [7b] &  8.574  &  0.522 &  0.044 & 0.06 & $-$0.03 &   ... &  $-$21.24  &  $-$25.68 \\
NGC 7768  & E2      & 112.8 [6h]& y [4l]& 257 [6h] &   13$^{+    5}_{-    4}$ [6h]  &  13.24      &  9.335  &  0.167 &  0.014 & 0.11 & $-$0.05 &  ...  &  $-$22.20  &  $-$26.11 \\
\multicolumn{15}{c}{45 Bulges} \\ 
Cygnus A  & Sa?     & 232.0 [3a]& y?    & 270      &   25$^{+    7}_{-    7}$ [6s]  &  17.04      & 10.276  &  1.644 &  0.140 & 0.17 & $-$0.07 &  0.13 &  $-$20.80  &  $-$25.81 \\
IC 2560   & SBb     &  40.7 [3a]&n? [4m]& 144 [5d] &0.044$^{+0.044}_{-0.022}$ [6t]  &  12.53 [7b] &  8.694  &  0.410 &  0.035 & 0.02 & $-$0.02 &  0.20 &  $-$19.02  &  $-$22.78 \\
NGC  224  & Sb      &   0.74    & n [4n]& 170      &  1.4$^{+  0.9}_{-  0.3}$ [6u]  &   4.36      &  0.984  &  0.268 &  0.023 & 0.00 & $-$0.00 &  0.49 &  $-$18.67  &  $-$21.82 \\ 
NGC  253  & SBc     &   3.5 [3c]& n [4o]& 109 [5e] & 0.10$^{+ 0.10}_{- 0.05}$ [6v]  &   8.04      &  3.772  &  0.081 &  0.007 & 0.00 & $-$0.00 &  0.61 &  $-$17.33  &  $-$21.42 \\
NGC  524  & S0      &  23.3     & y [4e]& 253      &  8.3$^{+  2.7}_{-  1.3}$ [6w]  &  11.30      &  7.163  &  0.356 &  0.030 & 0.03 & $-$0.02 &  0.01 &  $-$20.15  &  $-$23.61 \\
NGC 1023  & SB0     &  11.1     & n [4g]& 204      & 0.42$^{+ 0.04}_{- 0.04}$ [6x]  &  10.35      &  6.238  &  0.262 &  0.022 & 0.01 & $-$0.00 &  0.47 &  $-$19.82  &  $-$23.01 \\
NGC 1068  & SBb     &  15.2 [3a]& n [4p]& 165 [5f] &0.084$^{+0.003}_{-0.003}$ [6y]  &   9.61      &  5.788  &  0.145 &  0.012 & 0.01 & $-$0.01 &  0.07 &  $-$19.45  &  $-$23.47 \\
NGC 1194  & S0      &  53.9 [3a]& n?    & 148 [5g] & 0.66$^{+ 0.03}_{- 0.03}$ [6z]  &  13.83      &  9.758  &  0.330 &  0.028 & 0.05 & $-$0.03 &  0.25 &  $-$19.58  &  $-$22.91 \\
NGC 1300  & SBbc    &  20.7 [3a]& n [4q]& 229      & 0.73$^{+ 0.69}_{- 0.35}$ [6aa] &  11.11      &  7.564  &  0.130 &  0.011 & 0.01 & $-$0.01 &  0.18 &  $-$18.05  &  $-$21.91 \\
NGC 1316  & SB0     &  18.6 [3d]& ?     & 226      &  1.5$^{+ 0.75}_{-  0.8}$ [6ab] &   9.42      &  5.587  &  0.090 &  0.008 & 0.02 & $-$0.02 &  0.15 &  $-$21.33  &  $-$24.68 \\ 
NGC 1332  & S0      &  22.3     & y?    & 320      & 14.5$^{+    2}_{-    2}$ [6ac] &  11.25      &  7.052  &  0.141 &  0.012 & 0.02 & $-$0.01 &  0.51 &  $-$20.37  &  $-$23.74 \\
NGC 2273  & SBa     &  28.5 [3a]& n [4r]& 145 [5g] &0.083$^{+0.004}_{-0.004}$ [6z]  &  12.55      &  8.480  &  0.305 &  0.026 & 0.02 & $-$0.01 &  0.12 &  $-$19.33  &  $-$22.66 \\
NGC 2549  & SB0 [2a]&  12.3     & n [4q]& 144      & 0.14$^{+ 0.02}_{- 0.13}$ [6w]  &  12.19      &  8.046  &  0.282 &  0.024 & 0.01 & $-$0.01 &  0.48 &  $-$18.24  &  $-$21.44 \\
NGC 2778  & SB0 [2b]&  22.3     & n [4d]& 162      & 0.15$^{+ 0.09}_{-  0.1}$ [6c]  &  13.35      &  9.514  &  0.090 &  0.008 & 0.02 & $-$0.02 &  0.14 &  $-$17.79  &  $-$21.15 \\
NGC 2787  & SB0     &   7.3     & n [4s]& 210      & 0.40$^{+ 0.04}_{- 0.05}$ [6ad] &  11.82      &  7.263  &  0.565 &  0.048 & 0.01 & $-$0.00 &  0.19 &  $-$17.41  &  $-$20.98 \\
NGC 2960  & Sa?     &  81.0 [3e]& n?    & 166 [5g] &0.117$^{+0.005}_{-0.005}$ [6z]  &  13.29 [7b] &  9.783  &  0.193 &  0.016 & 0.05 & $-$0.03 &  0.16 &  $-$20.79  &  $-$23.69 \\
NGC 3031  & Sab     &   3.8     & n [4t]& 162      & 0.74$^{+ 0.21}_{- 0.11}$ [6ae] &   7.89      &  3.831  &  0.346 &  0.029 & 0.00 & $-$0.00 &  0.28 &  $-$19.59  &  $-$22.56 \\
NGC 3115  & S0      &   9.4     & n [4a]& 252      &  8.8$^{+ 10.0}_{-  2.7}$ [6af] &   9.87      &  5.883  &  0.205 &  0.017 & 0.01 & $-$0.01 &  0.47 &  $-$19.88  &  $-$23.00 \\ 
NGC 3227  & SBa     &  20.3 [3a]& n [4e]& 133      & 0.14$^{+ 0.10}_{- 0.06}$ [6ag] &  11.10      &  7.639  &  0.098 &  0.008 & 0.01 & $-$0.01 &  0.17 &  $-$19.87  &  $-$22.75 \\
NGC 3245  & S0      &  20.3     & n [4e]& 210      &  2.0$^{+  0.5}_{-  0.5}$ [6ah] &  11.70      &  7.862  &  0.108 &  0.009 & 0.02 & $-$0.01 &  0.26 &  $-$19.37  &  $-$22.61 \\
NGC 3368  & SBab    &  10.1     & n [4u]& 128      &0.073$^{+0.015}_{-0.015}$ [6ai] &  10.11      &  6.320  &  0.109 &  0.009 & 0.01 & $-$0.01 &  0.16 &  $-$19.14  &  $-$22.17 \\
NGC 3384  & SB0     &  11.3     & n [4a]& 148      & 0.17$^{+ 0.01}_{- 0.02}$ [6c]  &  10.85      &  6.750  &  0.115 &  0.010 & 0.01 & $-$0.01 &  0.34 &  $-$19.05  &  $-$22.47 \\
NGC 3393  & SBa     &  55.2 [3a]& n [4m]& 197      & 0.34$^{+ 0.02}_{- 0.02}$ [6aj] &  13.09 [7b] &  9.059  &  0.325 &  0.028 & 0.04 & $-$0.02 &  0.04 &  $-$20.21  &  $-$23.56 \\
NGC 3414  & S0      &  24.5     & n [4e]& 237      &  2.4$^{+  0.3}_{-  0.3}$ [6e]  &  11.96      &  7.981  &  0.106 &  0.009 & 0.02 & $-$0.01 &  0.14 &  $-$19.40  &  $-$22.87 \\
\enddata
\end{deluxetable*}

}

\setcounter{table}{0}

\scriptsize{
\begin{deluxetable*}{llllllllccccccc}
\tablecaption{{\it cont.}} 
\tablehead{ 
\colhead{Galaxy} & \colhead{Type} & \colhead{Dist} & \colhead{core} & \colhead{$\sigma$} & \colhead{$M_{\rm bh}$} & 
\colhead{$B_T$} & \colhead{$K_s$} & \colhead{A$_B$} & \colhead{A$_K$} & \colhead{K$_B$} & \colhead{K$_{K_s}$} & \colhead{R$_{25}$} 
& \colhead{M$_B$}  &  \colhead{M$_{K_s}$}  \\
\colhead{ } & \colhead{ } & \colhead{[Mpc]} & \colhead{ } & \colhead{km s$^{-1}$} & \colhead{[10$^8 M_{\odot}$]} & 
\colhead{[mag]} & \colhead{[mag]} & \colhead{[mag]} & \colhead{[mag]} & \colhead{[mag]} & \colhead{[mag]} & \colhead{ } 
&  \colhead{[mag]} & \colhead{[mag]}  \\
\colhead{(1)} & \colhead{(2)} & \colhead{(3)} & \colhead{(4)} & \colhead{(5)} & \colhead{(6)} & \colhead{(7)} & 
\colhead{(8)} & \colhead{(9)} & \colhead{(10)} & \colhead{(11)} & \colhead{(12)} & \colhead{(13)} & \colhead{(14)} & \colhead{(15)} 
}
\startdata 
NGC 3489  & SB0     &  11.7     & n [4q]& 105      &0.058$^{+0.008}_{-0.008}$ [6ai] &  11.12     &  7.370   &  0.072 &  0.006 & 0.01 & $-$0.01 &  0.24  &  $-$18.69 &  $-$21.88 \\
NGC 3585  & S0      &  19.5     & n [4q]& 206      &  3.1$^{+  1.4}_{-  0.6}$ [6o]  &  10.88     &  6.703   &  0.276 &  0.023 & 0.02 & $-$0.01 &  0.26  &  $-$20.27 &  $-$23.69 \\ 
NGC 3607  & S0      &  22.2     & n [4e]& 224      &  1.3$^{+  0.5}_{-  0.5}$ [6o]  &  10.82     &  6.994   &  0.090 &  0.008 & 0.01 & $-$0.01 &  0.30  &  $-$20.47 &  $-$23.68 \\
NGC 3998  & S0      &  13.7     &y? [4v]& 305      &  8.1$^{+  2.0}_{-  1.9}$ [6ak] &  11.61     &  7.365   &  0.069 &  0.006 & 0.01 & $-$0.01 &  0.08  &  $-$18.41 &  $-$22.21 \\ 
NGC 4026  & S0      &  13.2     & n [4q]& 178      &  1.8$^{+  0.6}_{-  0.3}$ [6o]  &  11.67     &  7.584   &  0.095 &  0.008 & 0.01 & $-$0.01 &  0.61  &  $-$18.88 &  $-$22.11 \\
NGC 4151  & SBab    &  20.0 [3a]& n [4e]& 156      & 0.65$^{+ 0.07}_{- 0.07}$ [6al] &  11.50     &  7.381   &  0.119 &  0.010 & 0.01 & $-$0.01 &  0.15  &  $-$19.24 &  $-$22.60 \\
NGC 4258  & SBbc    &   7.2 [3f]& n [4w]& 134      & 0.39$^{+ 0.01}_{- 0.01}$ [6am] &   9.10     &  5.464   &  0.069 &  0.006 & 0.00 & $-$0.00 &  0.41  &  $-$17.95 &  $-$21.76 \\
NGC 4342  & S0      &  23.0 [3g]& n [4x]& 253      &  4.5$^{+  2.3}_{-  1.5}$ [6an] &  13.41     &  9.023   &  0.088 &  0.008 & 0.01 & $-$0.01 &  0.33  &  $-$17.99 &  $-$21.73 \\
NGC 4388  & Sb      &  17.0 [3b]&n? [4y]& 107 [5g] &0.075$^{+0.002}_{-0.002}$ [6z]  &  11.76     &  8.004   &  0.143 &  0.012 & 0.02 & $-$0.02 &  0.64  &  $-$20.07 &  $-$23.65 \\
NGC 4459  & S0      &  15.7     & n [4f]& 178      & 0.68$^{+ 0.13}_{- 0.13}$ [6ad] &  11.32     &  7.152   &  0.199 &  0.017 & 0.01 & $-$0.01 &  0.12  &  $-$19.15 &  $-$22.73 \\
NGC 4564  & S0 [2c] &  14.6     & n [4f]& 157      & 0.60$^{+ 0.03}_{- 0.09}$ [6c]  &  12.05     &  7.937   &  0.151 &  0.013 & 0.01 & $-$0.01 &  0.38  &  $-$18.49 &  $-$21.87 \\
NGC 4594  & Sa      &   9.5     &y [6ao]& 297 [6ao]&  6.4$^{+  0.4}_{-  0.4}$ [6ao] &   8.98     &  4.962   &  0.221 &  0.019 & 0.01 & $-$0.01 &  0.39  &  $-$20.71 &  $-$23.86 \\ 
NGC 4596  & SB0     &  17.0 [3b]& n [4z]& 149      & 0.79$^{+ 0.38}_{- 0.33}$ [6ad] &  11.35     &  7.463   &  0.096 &  0.008 & 0.02 & $-$0.01 &  0.13  &  $-$19.20 &  $-$22.60 \\
NGC 4736  & Sab     &  4.4  [3i]& n?    & 104      &0.060$^{+0.014}_{-0.014}$ [6ap] &   8.99     &  5.106   &  0.076 &  0.007 & 0.00 & $-$0.00 &  0.09  &  $-$18.37 &  $-$21.55 \\
NGC 4826  & Sab     &  7.3      & n?    &  91      &0.016$^{+0.004}_{-0.004}$ [6ap] &   9.36     &  5.330   &  0.178 &  0.015 & 0.00 & $-$0.00 &  0.27  &  $-$19.36 &  $-$22.47 \\
NGC 4945  & SBcd    &   3.8 [3i]&n? [4aa]& 100     &0.014$^{+0.014}_{-0.007}$ [6aq] &   9.30     &  4.483   &  0.762 &  0.065 & 0.00 & $-$0.00 &  0.72  &  $-$16.39 &  $-$20.60 \\
NGC 5128  & S0      &   3.8 [3i]&n? [4ab]& 120     & 0.45$^{+ 0.17}_{- 0.10}$ [6ar] &   7.84     &  3.942   &  0.496 &  0.042 & 0.01 & $-$0.00 &  0.11  &  $-$19.84 &  $-$22.87 \\ 
NGC 6264  & Sa?     & 146.3 [3a]& n?    & 159 [5g] &0.305$^{+0.004}_{-0.004}$ [6z]  & 15.42 [7b] & 11.407   &  0.280 &  0.024 & 0.10 & $-$0.07 &  0.15  &  $-$20.04 &  $-$23.47 \\
NGC 6323  &SBab [2d]& 112.4 [3a]& n?    & 159 [5g] & 0.10$^{+0.001}_{-0.001}$ [6z]  & 14.85 [7b] & 10.530   &  0.074 &  0.006 & 0.06 & $-$0.05 &  0.48  &  $-$20.01 &  $-$23.44 \\
NGC 7582  & SBab    &  22.0 [3a]&n [4ac]& 156      & 0.55$^{+ 0.26}_{- 0.19}$ [6as] &  11.37     &  7.316   &  0.061 &  0.005 & 0.01 & $-$0.01 &  0.38  &  $-$19.77 &  $-$22.94 \\
UGC 3789  & SBab    &  48.4 [3a]& n?    & 107 [5g] &0.108$^{+0.005}_{-0.005}$ [6z]  & 13.30 [7b] &  9.510   &  0.280 &  0.024 & 0.03 & $-$0.04 &  0.06  &  $-$19.48 &  $-$22.46 \\
\multicolumn{15}{c}{3 extra} \\ %
NGC 3079  & SBc     & 20.7 [3a]&n? [4ad]&60-150 [5h]&0.024$^{+0.024}_{-0.012}$ [6at]&  11.54     &  7.262   &  0.049 &  0.004 & 0.01 & $-$0.00 &  0.74  &  $-$17.80 &  $-$21.86 \\
Milky Way & SBbc    &0.008 [6au]&n [4ae]& 100 [5i] &0.043$^{+0.004}_{-0.004}$ [6au] &    ...     &   ...    &    ... &    ... &  ... &  ...    &   ...  &  ...      &   ...     \\
M32       & E? [2e]   & 0.79      &n [4ae]& 55  [5j] &0.024$^{+0.005}_{-0.005}$ [6av] &   9.03     &  5.095   &  0.268 &  0.023 & 0.00 & $-$0.00 &  0.13  &  $-$15.73 &  $-$19.42 \\
\enddata
\tablecomments{
Column 1: Galaxy name. 
Column 2: Basic morphological type, primarily from NED\footnote{http://nedwww.ipac.caltech.edu}. 
Column 3: Distance, primarily from Tonry et al.\ (2001) and corrected according to Blakeslee et al.\ (2002). 
Column 4: Presence of a partially depleted core.  The addition of a question mark is used to indicate that this
classification has come from the velocity dispersion criteria mentioned in the text.
Column~5: Velocity dispersion primarily from HyperLeda\footnote{http://leda.univ-lyon1.fr} (Paturel et al.\ 2003). 
Column~6: Black hole mass, adjusted to the distances in column~3. 
Column~7:  $B_T$ is the total (observed) magnitude in the $B$ system from the RC3, unless otherwise noted.
Column~8: $K_s$ is the total 2MASS $K_s$-band magnitude. 
Columns~9~\&~10: The $B$ and $K$-band Galactic extinction from Schlegel et al.\ (1998). 
Columns~11~\&~12: The $B$ and $K$-band K-correction. 
Column~13: R$_{25}$, taken from the RC3, is the (base 10) logarithm of the
major-to-minor ($a/b$) isophotal axis ratio at $\mu_B$=25 mag arcsec$^{-2}$. 
Columns~14~\&~15: Corrected, absolute, $B$- and $K$-band spheroid magnitude. 
References: 
2a = Krajnovi\'c et al.\ (2009); 
2b = Rix et al.\ (1999); 
2c = Trujillo et al.\ (2004); 
2d = Jarrett et al.\ (2000) 2MASS image; 
2e = Graham (2002); 
3a = NED (Virgo + GA + Shapley)-corrected Hubble flow distance; 
3b = Jerjen et al.\ (2004); 
3c = Rekola et al.\ (2005); 
3d = Madore et al.\ (1999); 
3e = Violette Impellizzeri et al.\ (2012); 
3f = Herrnstein et al.\ (1999); 
3g = Mei et al.\ (2007), Blakeslee et al.\ (2009), Blom et al.\ (2012), Bogd\'an et al. (2012b); 
3h = Herrmann et al.\ (2008); 
3i = Karachentsev et al.\ (2007); 
4a = Ravindranath et al.\ (2001); 
4b = Dullo \& Graham (2012); 
4c = Lauer et al.\ (2005); 
4d = Rest et al.\ (2001); 
4e = Richings et al.\ (2011); 
4f = Ferrarese et al.\ (2006); 
4g = Faber et al.\ (1997); 
4h = Trujillo et al.\ (2004); 
4i = Forbes, Brodie, \& Huchra (1997);
4j = Laine et al.\ (2003); 
4k = Quillen et al.\ (2000); 
4l = Grillmair et al.\ (1994); 
4m = Mu{\~n}oz Mar{\'{\i}}n et al.\ (2007); 
4n = Corbin, O'Neil, \& Rieke (2001); 
4o = Kornei \& McCrady (2009); 
4p = Macchetto et al.\ (1994); 
4q = Graham (2012b); 
4r = Malkan, Gorjian \& Tam (1998); 
4s = Erwin et al.\ (2003); 
4t = Fisher \& Drory (2008); 
4u = Nowak et al.\ (2010); 
4v = Gonz\'alez Delgado et al.\ (2008); 
4w = Pastorini et al.\ (2007); 
4x = van den Bosch, Jaffe, \& van der Marel (1998); 
4y = Pogge \& Martini (2002); 
4z = Gerssen, Kuijken, \& Merrifield (2002); 
4aa = Marconi et al.\ (2000); 
4ab = Radomski et al.\ (2008); 
4ac = Wold \& Galliano (2006); 
4ad = Cecil et al.\ (2001);
4ae = Graham \& Spitler (2009); 
5a = Dalla Bont\`a et al.\ (2009); 
5b = Smith et al.\ (2000); 
5c = De Francesco et al.\ (2006); 
5d = Cid Fernandes et al.\ (2004); 
5e = Oliva et al.\ (1995); 
5f = Nelson \& Whittle (1995); 
5g = Greene et al.\ (2010); 
5h = Shaw, Wilkinson, \& Carter (1993); 
5i = Merritt \& Ferrarese (2001a);
5j = Lucey et al.\ (1997), Chilingarian (2012, in prep.); 
6a = Dalla Bont\`a et al.\ (2009); 
6b = Cappellari et al.\ (2002), stellar dynamical measurement; 
6c = Gebhardt et al.\ (2003), G\"ultekin et al.\ (2009b); 
6d = Houghton et al.\ (2006), Gebhardt et al.\ (2007); 
6e = Preliminary values determined by Hu (2008) from Conf.\ Proc.\ figures of Cappellari et al.\ (2008); 
6f = Copin et al.\ (2004); 
6g = van den Bosch \& de Zeeuw (2010); 
6h = McConnell et al.\ (2012); 
6i = Ferrarese et al.\ (1996); 
6j = Walsh et al.\ (2010); 
6k = Gebhardt et al.\ (2011); 
6l = Nowak et al.\ (2007); 
6m = Shen \& Gebhardt (2009); 
6n = De Francesco et al.\ (2008); 
6o = G\"ultekin et al.\ (2009a); 
6p = McConnell et al.\ (2011b); 
6q = Ferrarese \& Ford (1999); 
6r = van der Marel \& van den Bosch (1998); 
6s = Tadhunter et al.\ (2003); 
6t = Ishihara et al.\ (2001), Nakai et al.\ (1998); 
6u = Bacon et al.\ (2001), Bender et al.\ (2005); 
6v = Rodr{\'{\i}}guez-Rico et al.\ (2006), a factor of 2 uncertainty has been assigned here; 
6w = Krajnovi\'c et al.\ (2009); 
6x = Bower et al.\ (2001); 
6y = Lodato \& Bertin (2003); 
6z = Kuo et al.\ (2011); 
6aa = Atkinson et al.\ (2005); 
6ab = Nowak et al.\ (2008); 
6ac = Rusli et al.\ (2011); 
6ad = Sarzi et al.\ (2001); 
6ae = Devereux et al.\ (2003); 
6af = Emsellem et al.\ (1999); 
6ag = Davies et al.\ (2006), Hicks \& Malkan (2008); 
6ah = Barth et al.\ (2001); 
6ai = Nowak et al.\ (2010); 
6aj = Kondratko et al.\ (2008); 
6ak = Walsh et al.\ (2012); 
6al = Onken et al.\ (2007), Hicks \& Malkan (2008); 
6am = Herrnstein et al.\ (1999); 
6an = Cretton \& van den Bosch (1999), Valluri et al.\ (2004); 
6ao = Jardel et al.\ (2011), NGC~4594 is a core-S\'ersic galaxy with an AGN; 
6ap = Kormendy et al.\ (2011), priv.\ value from K.Gebhardt; 
6aq = Greenhill et al.\ (1997); 
6ar = Neumayer (2010); 
6as = Wold et al.\ (2006); 
6at = Trotter et al.\ (1998), Yamauchi et al.\ (2004), Kondratko et al.\ (2005); 
6au = Gillessen et al.\ (2009);
6av = Verolme et al.\ (2002);
%
%
7a = $B_T$ (Johnson), Gavazzi et al.\ (2005); 
7b = The RC3's $m_B$ value (which they reduced to the $B_T$ system); 
8a = Cappellari et al.\ (2006).
}
\end{deluxetable*}

}    
\normalsize{ }











\subsection{Bulge magnitudes}

In addition to the above three corrections that were applied in the analysis
rather than in Table~\ref{Tab1}, the tabulated disk galaxy magnitudes
were adjusted for two other factors: their {\it observed, total} 
magnitudes were converted into {\it dust-corrected, bulge} magnitudes.  Rather
than do this galaxy by galaxy (e.g.\ Grootes et al.\ 2012), which would 
require careful bulge/disk decompositions, we can take advantage of our large
sample size and employ a mean statistical correction.  While this will result
in individual bulge magnitudes not being exactly correct, the ensemble average
correction 
will be correct.  Given that dust has an order of magnitude less impact in the
$K_s$-band than in the $B$-band, we can expect there to be less scatter in our
$K_s$-band $M_{\rm bh}$--$L$ relation, at least for the S\'ersic sample which
contains the bulk of the bulges. 

Using the relation $M=-2.5\log(L)$, the observed and the intrinsic (i.e.\
dust-corrected) bulge and disk luminosities are related as follows.
\begin{equation}
L_{\rm bulge,obs}/L_{\rm bulge,int} = 10^{-\Delta M_{\rm bulge}/2.5}, \nonumber 
\end{equation}
\begin{equation}
L_{\rm disk,obs}/L_{\rm disk,int} = 10^{-\Delta M_{\rm disk}/2.5}, \nonumber 
\end{equation}
\begin{equation}
L_{\rm bulge,obs} + L_{\rm disk,obs} = 10^{-M_{\rm galaxy,obs}/2.5}, \nonumber 
\end{equation}
\begin{equation}
L_{\rm bulge,int}/L_{\rm disk,int} = B/D, \nonumber 
\end{equation}
where $\Delta M_{\rm bulge}$ and $\Delta M_{\rm disk}$ are the differences
between the observed and intrinsic magnitudes (given below).  The above 
equations can be combined to give the useful expression
\begin{eqnarray}
M_{\rm bulge,int} &=& M_{\rm galaxy,obs} \nonumber  \\
 & &  + 2.5\log\left[ 10^{-\Delta M_{\rm bulge}/2.5} +
\frac{10^{-\Delta M_{\rm disk}/2.5}}{(B/D)}\right].  
\end{eqnarray}
 
The average dust-corrected bulge-to-disk ($B/D$) flux ratio is well known to be
a function of both galaxy morphological type and passband, varying mainly due
to the bulge luminosity (e.g.\ Yoshizawa \& Wakamatsu 1975; Trujillo et al.\
2001).  
Graham \& Worley (2008) presented over 400 $K$-band, $B/D$ flux ratios as a function of disk
galaxy morphological type.  Given that extinction due to dust is still an
issue at near-infrared wavelengths, albeit notably less in the $K$-band than the 
$H$-band, these flux ratios were corrected by Graham \& Worley (2008) for dust following the
prescription given by Driver et al.\ (2008).  Most studies have not 
made this important correction which effectively increases the observed 
$B/D$ flux ratio after accounting for the light from the far-side of the bulge
which is obscured by the centrally-concentrated dust in the disk. 
Here we have slightly 
adjusted Graham \& Worley's (2008) $B/D$ values by combining the S0 and S0/a
classes as one in our Table~\ref{Tab2} and ensuring that 
they do not have a smaller $B/D$ ratio than the Sa galaxies. 
By combining these ratios with the average, inclination-dependent $\Delta
M_{\rm bulge}$ and $\Delta M_{\rm disk}$ values below, one can derive the
expected `intrinsic' bulge magnitude from the observed (dust-dimmed) galaxy
magnitude and the inclination of the disk.  
%
Specifically, for $\Delta M_{\rm bulge}$ and $\Delta M_{\rm disk}$ we have
used the expressions from Driver et al.\ (2008) such that 
\begin{equation} 
\Delta M_{\rm bulge} = M_{\rm bulge, obs} - M_{\rm bulge, intrin} =
b_1 + b_2 \left[ 1-\cos(i)\right] ^{b_3}, 
\label{Eq_Mb}
\end{equation}
and 
\begin{equation}
\Delta M_{\rm disk} = M_{\rm disc, obs} - M_{\rm disc, intrin}  =
d_1 + d_2 \left[ 1-\cos(i)\right] ^{d_3},
\label{Eq_Md}
\end{equation}
where the coefficients in the above equations depend on the passband used and
are provided in Driver et al.\ (2008). 
The cosine of the disk inclination angle $i$, such that $i=0$
degrees for a face-on disk, is roughly equal to the minor-to-major axis ratio,
$b/a$, at large radii. For our galaxy sample, these values have been provided
in the final column of Table~\ref{Tab1}.

\begin{table}
\begin{center}
\caption{Average dust-corrected (intrinsic) bulge-to-disk flux ratios} 
\label{Tab2}
\begin{tabular}{@{}lcc@{}}
\hline
Galaxy  &  $<$$\log(B/D)$$>$  &  $<$$\log(B/D)$$>$  \\
Type    &  $B$-band     &  K-band       \\
\hline  
S0/S0a  &   -0.29       &   -0.31       \\
Sa      &   -0.29       &   -0.34       \\
Sab     &   -0.39       &   -0.54       \\
Sb      &   -0.87       &   -0.60       \\
Sbc     &   -1.13       &   -0.82       \\
Sc      &   -1.28       &   -1.06       \\
Scd     &   -1.55       &   -1.23       \\
\hline
\end{tabular}

Adapted from Graham \& Worley (2008).
\end{center}
\end{table}

We have used a conservative (small) uncertainty of 5\% for the velocity dispersions, 
and assigned a typical uncertainty of 0.25 mag to the elliptical galaxy magnitudes.  
Regarding the disk galaxies, the observed range of bulge-to-disk flux ratios
(Graham \& Worley 2008) has a 1$\sigma$ scatter equal to a factor of $\sim$2
for any given disk galaxy morphological type.  We therefore assign a notably
larger uncertainty of 0.75 mag to our bulge magnitudes.
%
%
While this may at first sound worryingly high, we again note that the {\it 
sample average} shift from a disk galaxy magnitude to a bulge magnitude will be
more accurate, thereby much less affecting the recovery of the $M_{\rm
bh}$--$L$ relation.  It is only now that the disk galaxy sample size $N=44$ is
sufficiently large (coupled with 9 elliptical galaxies in the S\'ersic sample) 
that we can use this methodology because the uncertainty
in the disk-galaxy sample average correction scales with $1/\sqrt N$.

\subsection{Galaxy core type}\label{SecCT} 

As described in the Introduction, several global properties and scaling relations
of galaxies depend on their core type. 
We have therefore identified galaxies with or without a partially 
depleted stellar core.  Such cores, with shallow light profiles, are thought to have formed during `dry'
galaxy merger events in which a binary supermassive black hole's orbit decays
by gravitationally ejecting stars from the center of the newly merged 
galaxy (e.g.\ Begelman et al.\ 1980; Ebisuzaki et al.\ 
1991; Graham 2004; Merritt et al.\ 2007). 
%
However S\'ersic galaxies that have $n \lesssim$ 2--3 may also have a resolved
inner, negative logarithmic light-profile slope (for the spheroid) which is
less steep than 0.3, although they have no depleted core relative to their
outer profile.  Graham et al.\ (2003) warned that some of these galaxies 
may be misidentified as `core' galaxies and Dullo \& Graham (2012) have shown
that this has occurred $\sim$20\% of the time in the past. 
We therefore have a preference for cores identified as a deficit relative to the
inward extrapolation of the spheroid's outer S\'ersic profile (e.g.\ Trujillo
et al.\ 2004; Ferrarese et al.\ 2006; Dullo \& Graham 2012).  Such
galaxies are referred to as `core-S\'ersic' galaxies, while those with no
deficit or instead additional nuclear components are referred to as `S\'ersic' galaxies 
(e.g.\ Graham et al.\ 2003; Graham \& Guzm\'an 2003; Balcells et al.\
2003; Trujillo et al.\ 2004). 

For most of the luminous galaxies we have been able to identify from suitably
high-resolution images whether or not they possess a partially-depleted core,
although some are at too great a distance to know.  At the low-luminosity end
of the relation, dusty nuclei can make this task more challenging. When no
core designation was available or possible from the literature (see column~4
of Table~\ref{Tab1}) we used the velocity dispersion to help us assign the
core type.  This meant that for seven galaxies with $\sigma > 270$ km s$^{-1}$,
all but one of which actually have $\sigma > 300$ km s$^{-1}$, we classified them as
core-S\'ersic galaxies.  For 12 galaxies with $\sigma \le 166$ km
s$^{-1}$ we have classified them as S\'ersic galaxies.  This approach relates
to the trend in which low-luminosity spheroids, often found in disk galaxies,
do not possess partially depleted cores while luminous spheroids do (e.g.\
Nieto et al.\ 1991; Ferrarese et al.\ 1994; Faber et al.\ 1997). 
For the remaining galaxies, their core type was determined from
high-resolution images (see column~4 of Table~\ref{Tab1}). 


%
While Faber et al.\ (1997) reported that the peculiar galaxy NGC~1316
(Fornax~A) has a partially-depleted core, we feel that a proper bulge/disk/bar
decomposition --- which is beyond the scope of this paper --- might be 
required for this barred S0 galaxy. 
This may reveal a bulge with a small S\'ersic index and a shallow
inner profile slope, rather than an actual depleted core.  Although NGC~1316
is a merger remnant (Deshmukh et al.\ 2012, and references therein) --- which 
may support the presence of a depleted core if it had been a dry merger event --- 
the presence of much dust and gas, plus the existence of a bar, is 
indicative of a minor, wet merger.  
We note that no other {\it barred} galaxy in our sample has a core which is 
partially depleted of stars, and to be safe we felt it best to exclude this 
galaxy from our final analysis until a more detailed investigation of its stellar 
distribution is performed.
%


\section{Analysis and Results}\label{SecAaR}



\begin{figure*}
\includegraphics[angle=-90,scale=0.91]{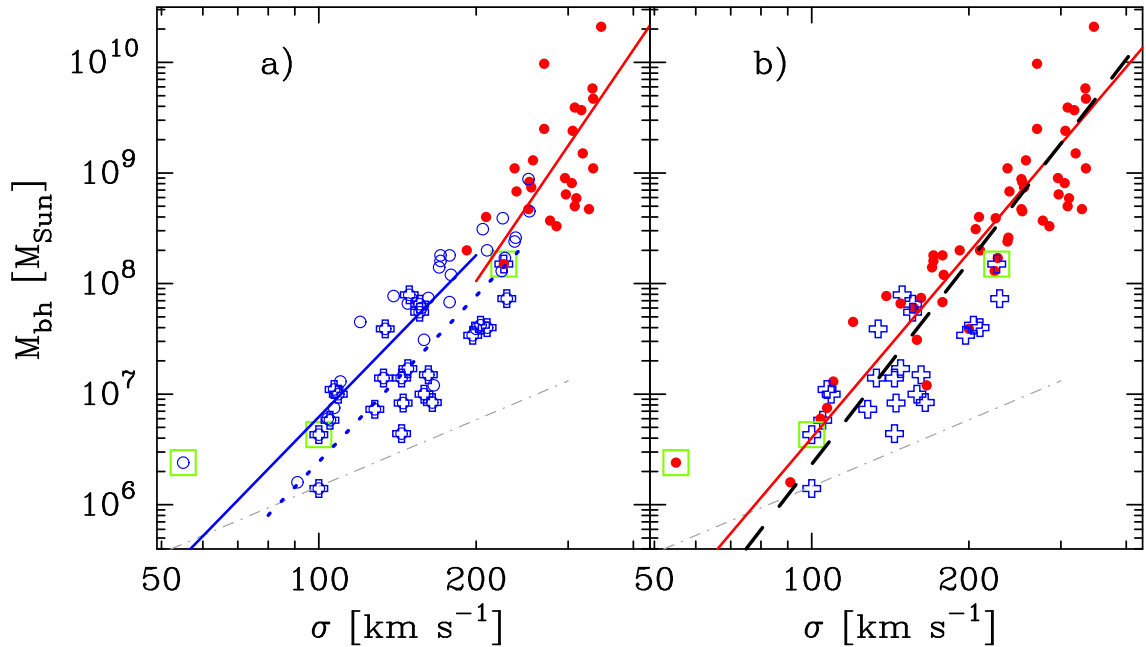}
\caption{
Black hole mass versus central velocity dispersion.  
The faint dot-dashed gray line corresponds to a black hole's
sphere-of-influence of $0\arcsec.1$ at a distance of 1.3 Mpc.  With such a
limiting spatial resolution of $0\arcsec.1$, black holes residing below this
line cannot be reliably detected.  
Panel a)  Red dots represent the core-S\'ersic galaxies, open blue circles represent
the S\'ersic galaxies, while the open 
crosses designate those which are barred.  All 77 galaxies shown here have 
been included in the analysis of this Figure.  
The upper solid red line corresponds to the
symmetrical regression for the 27 core-S\'ersic galaxies, while (due to the
sample selection limit) the dotted and solid blue lines correspond to the
`ordinary least squares' OLS($\sigma | M_{\rm bh}$) regression for the barred
and non-barred S\'ersic galaxies (see Table~\ref{Tab3}).
Panel b)  Red dots represent the non-barred galaxies while the open
blue crosses designate those which are barred.  The solid red line is a fit to
the non-barred galaxies, while the dashed black line is a fit to all
galaxies. 
M32, the Milky Way and Fornax~A (NGC~1316)
have been marked with a green square, they are not included in
Figure~\ref{Fig2} for the reasons mentioned in Section~\ref{Sec_Data}. 
The S\'ersic galaxy with the highest black hole mass is NGC~3115 (see
section~\ref{SecCT}). 
}
\label{Fig1}
\end{figure*}

Our initial investigation of the $M_{\rm bh}$--$\sigma$ diagram uses all 75
galaxies (including NGC~1316) plus M32 and the Milky Way's bulge to give a
total sample of 77 galaxies.  It is apparent that there is something
unusual with the two ultramassive black holes reported for NGC~3842 and
NGC~4889 by McConnell et al.\ (2011a). 
Figure~\ref{Fig1} reveals that they may be 
are outliers biasing the $M_{\rm bh}$--$\sigma$ relation defined by the
remaining data. 
Performing a symmetrical linear regression on the core-S\'ersic galaxies,
including NGC~3842 and NGC~4889, yields a slope of roughly 7.0$\pm$1 (see
Table~\ref{Tab3}).  
Uncertain as to whether there is an error in the black hole masses for
NGC~3842 and NGC~4889, or perhaps with their velocity dispersion measurements,
or if instead these values are fine and these 
galaxies are representative (of a possibly new class) of galaxies which
deviate from the $M_{\rm bh}$--$\sigma$ relation as we know it, in the 
following section we have elected to repeat the $M_{\rm bh}$--$\sigma$
regression analysis without these two galaxies.  If these galaxies are
outliers, then this additional approach provides a more robust statistical
analysis of the $M_{\rm bh}$--$\sigma$ relation.  To avoid issues of
inconsistent samples, these two galaxies are also excluded from the ensuing
$M_{\rm bh}$--$L$ diagram, although doing so has no significant effect on our 
$M_{\rm bh}$--$L$ relation.

Although NGC~1316 is not an outlier from the $M_{\rm bh}$--$\sigma$ diagram
(Figure~\ref{Fig1}), we do additionally exclude it from what follows due to
the previously mentioned uncertainty associated with its core type.  If
NGC~1316 is confirmed to possess a partially-depleted core, something
associated with dry merger events, then it will be the only such example in a
barred galaxy (with bars of course associated with secular evolution rather
than dry mergers).  Until a three-component S\'ersic-bulge plus
exponential-disk plus bar model has been fit to this galaxy, it is premature
to claim from a fitted single-component, core-S\'ersic model that this galaxy
has a depleted core.

Our cautious exclusion of these three galaxies (NGC~1316, NGC~3842 and
NGC~4889), leaves us with our final sample of (75-3=) 72 galaxies, comprised
of 24 core-S\'ersic galaxies and 48 S\'ersic galaxies.

\begin{table*}
\begin{center}
\caption{Black hole scaling relations} 
\label{Tab3}
\begin{tabular}{@{}rllccc@{}}
\hline
\multicolumn{2}{l}{\# and Type}  &  Regression &   $\alpha$    &   $\beta$       &  $\Delta$ dex \\
\hline
\multicolumn{6}{c}{Figure~\ref{Fig1} (includes some questionable data points)}\\
\multicolumn{6}{c}{
$\log[M_{\rm bh}/M_{\odot}] = \alpha + \beta \log[\sigma/200\, {\rm km} s^{-1}]$}  \vspace{1mm} \\
27 & Core-S\'ersic       & Bisector & $8.02\pm0.17$  &  $6.95\pm1.17$  & 0.45 \\
28 & Non-barred S\'ersic & Bisector & $8.21\pm0.06$  &  $4.39\pm0.53$  & 0.32 \\
23 & Barred              & Bisector & $7.78\pm0.10$  &  $4.14\pm0.55$  & 0.34 \\
54 & Non-barred All      & Bisector & $8.29\pm0.05$  &  $5.09\pm0.44$  & 0.39  \vspace{1mm} \\
28 & Non-barred S\'ersic & OLS($\sigma|M_{\rm bh}$) & $8.25\pm0.06$  &  $4.82\pm0.64$  & 0.34 \\
23 & Barred              & OLS($\sigma|M_{\rm bh}$) & $7.89\pm0.16$  &  $4.98\pm1.00$  & 0.37 \\
54 & Non-barred All      & OLS($\sigma|M_{\rm bh}$) & $8.28\pm0.06$  &  $5.57\pm0.52$  & 0.41 \\
77 & All                 & OLS($\sigma|M_{\rm bh}$) & $8.20\pm0.05$  &  $6.08\pm0.41$  & 0.46 \\
\hline 
\multicolumn{6}{c}{Figure~\ref{Fig2}, panel a), $M_{\rm bh}$--$\sigma$} \\
\multicolumn{6}{c}{
$\log[M_{\rm bh}/M_{\odot}] = \alpha + \beta \log[\sigma/200\, {\rm km} s^{-1}]$} \vspace{1mm} \\
24 & Core-S\'ersic       & Bisector &  $8.20\pm0.13$  &  $5.35\pm0.92$  & 0.35 \\
27 & Non-barred S\'ersic & Bisector &  $8.23\pm0.06$  &  $5.02\pm0.41$  & 0.31 \\
21 & Barred S\'ersic     & Bisector &  $7.75\pm0.12$  &  $4.03\pm0.69$  & 0.34 \\
51 & Non-barred All      & Bisector &  $8.23\pm0.05$  &  $5.14\pm0.31$  & 0.33  \\ 
28 & Elliptical          & Bisector &  $8.23\pm0.05$  &  $4.86\pm0.50$  & 0.34  \vspace{1mm} \\ 
27 & Non-barred S\'ersic & OLS($\sigma|M_{\rm bh}$) &  $8.27\pm0.06$  &  $5.53\pm0.51$  & 0.33 \\
{\bf 21} & {\bf Barred S\'ersic}   & {\bf OLS($\sigma|M_{\rm bh}$)} & {\bf 7.92$\pm$0.23} & {\bf 5.29$\pm$1.47} & {\bf 0.40} \\
{\bf 51} & {\bf Non-barred All} & {\bf OLS($\sigma|M_{\rm bh}$)} & {\bf 8.22$\pm$0.05} & {\bf 5.53$\pm$0.34} & {\bf 0.34} \\
28 & Elliptical          & OLS($\sigma|M_{\rm bh}$) &  $8.21\pm0.10$  &  $5.54\pm0.71$  & 0.37 \\
72 & All                 & OLS($\sigma|M_{\rm bh}$) &  $8.15\pm0.05$  &  $6.08\pm0.31$  & 0.41 \\
\hline 
\multicolumn{6}{c}{Figure~\ref{Fig2}, panel b), $M_{\rm bh}$--$L_{K_s,sph}$}\\
\multicolumn{6}{c}{$\log [M_{\rm bh}/M_{\odot}] = \alpha + \beta[M_{K_s,sph}+25]$} \vspace{1mm} \\
{\bf 24} & {\bf Core-S\'ersic}  & {\bf Bisector} &  {\bf 9.05$\pm$0.09}  & {\bf $-$0.44$\pm$0.08} & {\bf 0.44} \\
\multicolumn{6}{c}{$\log [M_{\rm bh}/M_{\odot}] = \alpha + \beta[M_{K_s,sph} +22.5]$} \\
{\bf 48} & {\bf S\'ersic}  & {\bf Bisector}  &  {\bf 7.39$\pm$0.14}  &  {\bf  $-$1.09$\pm$0.22} & {\bf 0.95} \\
\hline 
\multicolumn{6}{c}{Figure~\ref{Fig2}, panel c), $M_{\rm bh}$--$L_{B,sph}$}\\
\multicolumn{6}{c}{$\log [M_{\rm bh}/M_{\odot}] = \alpha + \beta[M_{B,sph} +21]$} \\
{\bf 24} & {\bf Core-S\'ersic}  & {\bf Bisector} &  {\bf 9.03$\pm$0.09}  & {\bf $-$0.54$\pm$0.12} & {\bf 0.44} \\
\multicolumn{6}{c}{$\log [M_{\rm bh}/M_{\odot}] = \alpha + \beta[M_{B,sph} +19]$} \\
{\bf 48} & {\bf S\'ersic}  & {\bf Bisector} &  {\bf 7.37$\pm$0.15}  &  {\bf $-$0.94$\pm$0.16} & {\bf 0.93} \\
\hline
\end{tabular}

$M_{\rm bh}$ = black hole mass, $\sigma$ = host galaxy velocity dispersion,
$M_{K_s,sph}$ = dust-corrected, 2MASS $K_s$-band spheroid magnitude,
$M_{B,sph}$ = dust-corrected, RC3 $B$-band spheroid magnitude.  A symmetrical
``bisector'' regression was used, in addition to a non-symmetrical `ordinary least squares'
OLS$(X|M_{\rm bh})$ regression to compensate for the sample selection limit,
or floor in the data, at 
the low-mass end (see Figure~\ref{Fig2}).  The total rms scatter in the $\log
M_{\rm bh}$ direction is given by $\Delta$, and roughly scales linearly with
the slope (within each panel). 
The lines highlighted in bold show the preferred fits, see the text for 
details, while Table~\ref{Tab4} provides revised $M_{\rm bh}$--$L_{K_s}$ 
relations according to Schombert \& Smith's (2012) correction of the 2MASS photometry.
\end{center}
\end{table*}

\subsection{The $M_{\rm bh}$--$\sigma$ diagram}\label{SecMs}

%
In Figure~\ref{Fig2}a we have performed a symmetrical linear regression for the
24 core-S\'ersic galaxies in the $M_{\rm bh}$--$\sigma$ diagram. 
This was achieved
using the bisector regression analysis in the {\sc BCES} routine from Akritas
\& Bershady (1996).  The results are provided in Table~\ref{Tab3}.  Of note is
that the symmetrical $M_{\rm bh}$--$\sigma$ relation has a slope consistent
with a value of 5 (Ferrarese \& Merritt 2000; Hu 2008; Graham et al.\ 2011).

We have additionally performed the same symmetrical regression for the
S\'ersic sample (see Table~\ref{Tab3}).  However, due to each supermassive
black hole's limited sphere-of-influence, $r_{\rm infl} \sim G M_{\rm
bh}/\sigma^2$ (Merritt \& Ferrarese 2001b), there is an associated galaxy
distance limit to which we can spatially resolve the black hole's dynamical
influence on its surrounding gas and stars. Shown in Figure~\ref{Fig2}a is the
limit to which one can detect black holes at a distance of 1.3 Mpc when the
spatial resolution is $0\arcsec.1$ (i.e.\ roughly that achievable with the
Hubble Space Telescope and ground-based adaptive-optics-assisted
observations).  For galaxies further away, this limit moves vertically in the
diagram to higher black hole masses.  Attempts to occupy the lower right of
Figure~\ref{Fig2}a may need to wait for the next generation of telescopes with
greater spatial resolution, and as such we currently have a selection limit, a
kind of floor, in this diagram (Batcheldor 2010; Schulze \&
Wisotzki 2011).  Galaxies can exist below this floor, but 
they are at distances such that we can not resolve their black hole's 
sphere-of-influence. 
To avoid this sample selection bias on the samples which extend into the lower
portion of the diagram, we have additionally performed an ordinary least
squares (OLS) regression of the abscissa (i.e., the horizontal $X$ value) on
the ordinate (i.e., the vertical $Y$ value) for the reasons expounded in
Lynden-Bell et al.\ (1988, their Figure~10).

Given that the 27 non-barred S\'ersic galaxies define an $M_{\rm
bh}$--$\sigma$ relation which is consistent with that defined by the 24
core-S\'ersic galaxies (see Table~\ref{Tab3}), they have been combined to
produce a single non-barred $M_{\rm bh}$--$\sigma$ relation.  These 51
galaxies yield a slope of 5.53$\pm$0.34, consistent with the value of
5.32$\pm$0.49 reported by Graham et al.\ (2011) using a sample of 44
non-barred galaxies. 
Barth, Greene, \& Ho (2005, their figure~2) offers further support for this
linear relation after reducing their active galactic nuclei (AGN) 
black hole mass estimates by a factor of $\sim$2 
due to a revision in the virial $f$-factor used to determine black hole masses
in AGN (Onken et al.\ 2004; Graham et al.\ 2011).
While Graham et al.\ (2011) discussed why a 5\% uncertainty on the velocity
dispersion may be optimistic, we note that using a 10\% uncertainty on
$\sigma$, rather than 5\%, results in the same slope to the $M_{\rm
bh}$--$\sigma$ relation when using the OLS($\sigma|M_{\rm bh}$) regression.
%
Coupling the knowledge that $L \propto \sigma^2$ for the S\'ersic spheroids, 
with the relation $M_{\rm bh} \propto \sigma^{5.5}$, one would expect to find 
$M_{\rm bh} \propto L^{2.75}$ for the S\'ersic spheroids. 

For a value of $\sigma = 400$ km s$^{-1}$, the non-barred $M_{\rm bh}$--$\sigma$
relation yields a black hole mass of $7.7 \times 10^9 M_{\odot}$.  This is 3
times greater than the value of $2.6\times10^9 M_{\odot}$ predicted by the
elliptical galaxy $M_{\rm bh}$--$\sigma$ relation from G\"ultekin et al.\
(2009b), and better, although not fully, 
matches the expectations from Hlavacek-Larrondo et al.\
(2012) for black holes in brightest cluster galaxies if they reside on the
``fundamental plane of black hole activity'' 
(Merloni et al.\ 2003; Falcke et al.\ 2004). 

As reported by Graham (2007b, 
2008a,b) and Hu (2008), the barred galaxies are, on average, offset to 
lower black hole masses or higher velocity dispersions than the non-barred 
galaxies, with the latter scenario possibly expected from barred galaxy dynamics 
(e.g.\ Gadotti \& Kauffmann 2009; Graham et al.\ 2011; DeBuhr, Ma \& White 2012). 
Consistent with Graham et 
al.\ (2011), we find a mean vertical offset of 0.30 dex between the barred and
non-barred galaxies in the $M_{\rm bh}$--$\sigma$ diagram, which corresponds to a factor of
$\sim$2 in black hole mass, when $\sigma = 200$ km s$^{-1}$. 


For those who may not know if their galaxy of interest is barred or not, using
the full 72 galaxies and an OLS($M_{\rm bh} | \sigma$), bisector, and
OLS($\sigma | M_{\rm bh}$) regression to construct the classical (all galaxy
type) $M_{\rm bh}$--$\sigma$ relation gives slopes of $5.21\pm0.27,
5.61\pm0.27$ and $6.08\pm0.31$, respectively.  The associated intercepts are
$8.14\pm0.04, 8.14\pm0.05$ and $8.15\pm0.05$.  These steeper relations are
expected given the offset nature of the barred galaxies.

\subsection{The $M_{\rm bh}$--$L$ diagram}

\begin{figure*}
\includegraphics[angle=-90,scale=0.75]{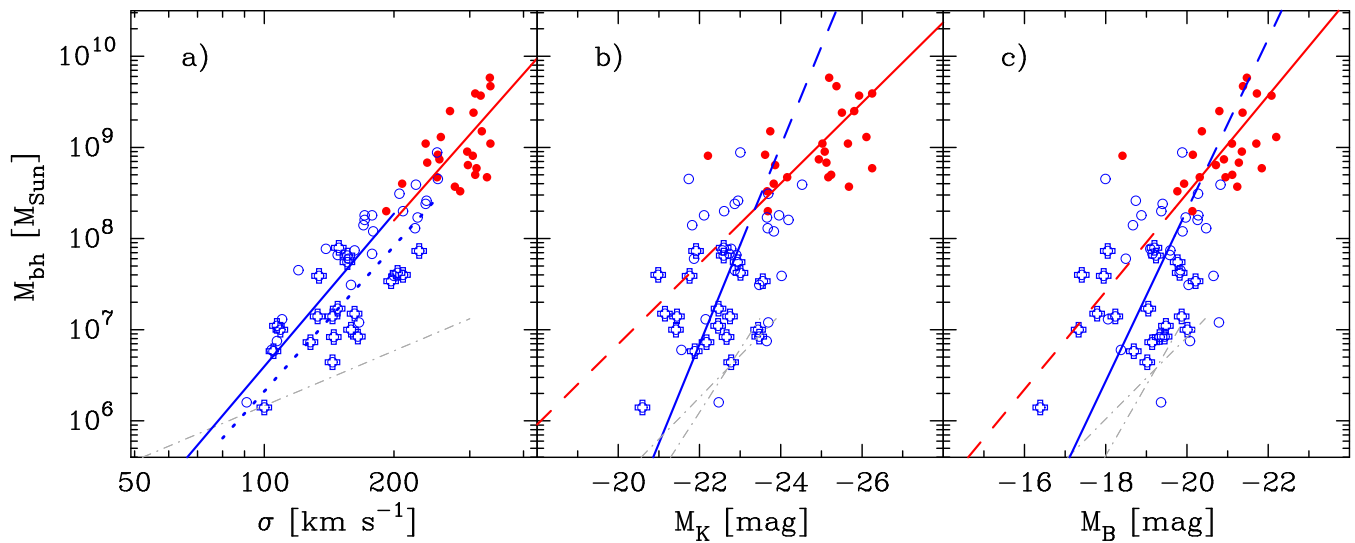}
\caption{
Black hole scaling relations involving a) central velocity dispersion
$\sigma$, b) $K_s$-band host spheroid magnitude and c) $B$-band host spheroid
magnitude.  The red dots represent the (luminous) core-S\'ersic galaxies and
the blue circles represent the (intermediate-luminosity) S\'ersic galaxies and
bulges, while the open crosses designate those which are barred.  
The symmetrical bisector regressions given
in Table~\ref{Tab3} are shown here, except for panel a) which shows the
OLS($\sigma|M_{\rm bh}$) regression for the 21 barred (dotted blue) and 27 non-barred (solid blue)
S\'ersic galaxies, along with the symmetrical bisector regression for the 24
core-S\'ersic galaxies. 
The faint dot-dashed gray line at the bottom of panel a) corresponds to a
black hole's sphere-of-influence of $0\arcsec.1$ at a distance of 1.3 Mpc.
With such a limiting spatial resolution of $0\arcsec.1$, black holes residing below
this line cannot be reliably detected.  Using the $L_{K_s}$--$\sigma$
and $L_B$--$\sigma$ relations for the S\'ersic galaxies, constructed using 
both a bisector regression and an OLS($L|\sigma$) 
regression, this gray line has been mapped into panels b) and c).
The left-most core-S\'ersic galaxy in panels b) and c) is NGC~3998. 
}
\label{Fig2}
\end{figure*}

In Figures~\ref{Fig2}b and \ref{Fig2}c we have performed a symmetrical linear
regression for the 24 core-S\'ersic galaxies in the $M_{\rm bh}$--$L$ diagram.
Expressing their spheroid magnitudes as luminosities, we 
find that $M_{\rm bh} \propto L_{K_s}^{1.10\pm0.20}$ and $M_{\rm bh} \propto
L_{B}^{1.35\pm0.30}$, in reasonable agreement with the relation $M_{\rm bh}
\propto L^{1.0}$ and with past analyses of predominantly bright
galaxies\footnote{The inclusion of NGC~3842 and NGC~4889 does not
  significantly change these results.}. 
%
We note that depending on the progenitor galaxy mass ratios --- among the
individual dry merger events which built one's core-S\'ersic sample --- the
slope of the $M_{\rm bh}$--$L$ relation can be expected to deviate slightly
from a value of 1.  Although, for galaxies built from a sufficient number of 
dry mergers, this should become a second order effect in terms of the 
overall evolutionary scenario in the $M_{\rm bh}$--$L$ diagram, for the 
reason described by Peng (2007) and Jahnke \& Maccio (2011). 

The effective sample selection boundary shown in Figure~\ref{Fig2}a has been
mapped into Figures~\ref{Fig2}b and \ref{Fig2}c by converting this line's
velocity dispersion into the {\it expected} magnitude according to the
$L$--$\sigma$ relations defined by the S\'ersic galaxies in these diagrams.
The need for an OLS($L|M_{\rm bh}$) regression on the S\'ersic galaxies in the $M_{\rm
bh}$--$L$ diagrams, rather than a symmetrical regression, is no longer
as apparent because the `floor' to the sample selection has significantly
shifted/rotated. 


It is not clear if the barred S\'ersic galaxies are offset 
from the non-barred S\'ersic galaxies in Figures~\ref{Fig2}b and \ref{Fig2}c
and so we have therefore grouped them together in our analysis.  The results
of which, given in Table~\ref{Tab3}, are such that $M_{\rm bh} \propto
L_{K_s}^{2.73\pm0.55}$ and $M_{\rm bh} \propto L_{B}^{2.35\pm0.40}$ 
for the S\'ersic spheroids\footnote{The inclusion of NGC~3079 does not change
  these results.}. 
We note in passing that it is expected that the color-magnitude relation for
S\'ersic spheroids (e.g., Tremonti et al.\ 2004; Jim\'enez et al.\ 2011), as
opposed to the flat color-magnitude relation for core-S\'ersic galaxies (as
discussed in Graham 2012c), will result in these two exponents not being equal
to each other.  However the current uncertainty on these two exponents does not
allow us to detect this difference. 

Finally, Schombert (2011) and Schombert \& Smith (2012, see their Appendix) have
recently detailed a photometry error in the 2MASS Extended Source Catalog (Jarrett et
al.\ 2000), such that the total 2MASS magnitudes are, on average, 0.33 mag too
faint in the $J$-band.
There is a similar offset in the $K_s$-band data (Schombert 2012,
priv.\ comm.) such that the $V-K_s$ colors are too blue by this 
amount. Therefore, for those using $K$ or $K_s$-band data not from the 2MASS
catalog for the prediction of black hole masses, the normalizing values of 25
and 22.5 used in the $M_{\rm bh}$--$L_{K_s}$ equations shown in
Table~\ref{Tab3} should be increased by 0.33 to 25.33 and 22.83.
One can actually go a little further than this. 
Assigning a 1-sigma uncertainty of 0.33/2 = 0.165 mag to this correction, that is,
assuming that the scatter seen in Schombert \& Smith's figure~12 has a
1-sigma value of 0.33/2 mag, we can estimate the impact of this correction on our
$K_s$-band $M_{\rm bh}$--$L$ relations.  Because they were constructed using
$N=24$ core-S\'ersic galaxies and $N=48$ S\'ersic galaxies, our `normalizing'
values of 25 and 22.5 mag have an additional uncertainty of $0.165/\sqrt{24}$
and $0.165/\sqrt{48}$ mag associated with them, which is a small but non-zero 
0.034 and 0.024 mag, respectively.  Given the slopes of our two $M_{\rm bh}$--$L_{K_s}$
relations, this is equivalent to increasing the uncertainty on their intercept 
from 0.09 to 0.12 and from 0.14 to 0.16, respectively.  The result of
applying these corrections is shown in Table~\ref{Tab4}). 
While this is a rather minor adjustment to our equations, it remains true that
those using the 2MASS catalog of $K_s$-band magnitudes for {\it individual} galaxies
may have the true galaxy magnitude wrong by the extent shown by Schombert \& Smith
(2012), which will thus affect their predicted black hole mass. 

It may be worth providing one additional comment in regard to photometric
errors. If the 2MASS magnitudes are shown to contain a systematic error such
that a greater fraction of galaxy light at large radii 
is increasingly missed in the intrinsically brighter galaxies, i.e. those with extended
light profiles, this will not explain the bend in the $M_{\rm bh}$--$L$ 
relation.  Bulges in (often truncated) exponential disk are not significantly
affected by this potential problem.  The affect is greater in galaxies with
higher S\'ersic indices, specifically 
the core-S\'ersic elliptical galaxies.  Such a correction would therefore act
to slightly reduce the slope of the core-S\'ersic $M_{\rm bh}$--$L$ relation,
thus increasing the apparent `bend' in the $M_{\rm bh}$--$L$ diagram between
the S\'ersic and the core-S\'ersic spheroids.

\begin{table}
\begin{center}
\caption{Corrected $M_{\rm bh}$--$L_{K_s}$ scaling relations} 
\label{Tab4}   
\begin{tabular}{@{}rllccc@{}}
\hline
\multicolumn{2}{l}{\# and Type}  &  Regression &   $\alpha$    &   $\beta$ &  $\Delta$ dex \\
\hline
\multicolumn{5}{c}{$\log [M_{\rm bh}/M_{\odot}] = \alpha + \beta[M_{K_s,sph} +25.33]$} \\
24 & Core-S\'ersic  & Bisector     &  $9.05\pm0.12$  &  $-0.44\pm0.08$  & 0.44 \\
\hline 
\multicolumn{5}{c}{$\log [M_{\rm bh}/M_{\odot}] = \alpha + \beta[M_{K_s,sph} +22.83]$} \\
48 & S\'ersic       & Bisector     &  $7.39\pm0.16$  &  $-1.09\pm0.22$  & 0.95 \\
\hline 
\end{tabular}

$K_s$-band $M_{\rm bh}$--$L$ relations after correcting for the 2MASS
magnitude errors noted by Schombert \& Smith (2012). 
\end{center}
\end{table}

\subsection{Propagation of errors}

It is important to appreciate the uncertainties on the slope and
intercept of the relations given in Tables~\ref{Tab3} and \ref{Tab4}, plus the
associated intrinsic scatter $\epsilon$.   For example, 
the maximum 1-sigma uncertainty on the predicted value of $M_{\rm bh}$ 
using the $M_{\rm bh}$--$L_{K_s}$ relation 
is acquired by assuming uncorrelated 
errors on the magnitude $M_{K_s}$ and the slope and intercept of the 
relation.  Gaussian error propagation for the linear equation
$y=(b\pm\delta b)(x\pm \delta x) + (a\pm \delta a)$, gives an
error on $y$ equal to 
\begin{eqnarray}
\delta y & = & \sqrt{ (dy/db)^2(\delta b)^2 + (dy/da)^2(\delta a)^2 +
 (dy/dx)^2(\delta x)^2 } \nonumber \\
 & = & \sqrt{ x^2(\delta b)^2 + (\delta a)^2 + b^2(\delta x)^2 }. \nonumber
\end{eqnarray}
In the presence of intrinsic scatter ($\epsilon$) in the $y$-direction, the uncertainty on $y$ is
\begin{eqnarray}
\delta y = \sqrt{ x^2(\delta b)^2 + (\delta a)^2 + b^2(\delta x)^2 +
  \epsilon^2}. \nonumber
\end{eqnarray}

For the S\'ersic $M_{\rm bh}$--$L_{K_s}$ relation in Table~\ref{Tab4}, 
$x=[M_{K_s,sph} +22.83]$, and thus 
\begin{eqnarray}
(\delta \log M_{\rm bh}/M_{\odot})^2 & = &
 [M_{K_s,sph} +22.83]^2(0.22)^2 + (0.16)^2  \nonumber \\ 
 & & \hskip-40pt +  (-1.09)^2 [\delta M_{K_s,sph}]^2 + \epsilon^2, \nonumber
\label{EqMerr}
\end{eqnarray}
where $\delta M_{K_s,sph}$ is the uncertainty associated with the spheroid's
magnitude. 

For the preferred non-barred $M_{\rm bh}$--$\sigma$ relation in Table~\ref{Tab3}, 
$x=\log(\sigma/200\, {\rm km\, s}^{-1})$, so $dx/d\sigma = 1/[\ln(10)\sigma]$,
and therefore 
\begin{eqnarray}
(\delta \log M_{\rm bh}/M_{\odot})^2 & = &
 [\log(\sigma/200\, {\rm km\, s}^{-1})]^2(0.34)^2 + (0.05)^2 \nonumber \\
 & & \hskip-40pt + [5.53/\ln(10)]^2 [\delta \sigma/\sigma]^2 + (\epsilon)^2. \nonumber
\label{EqM2e}
\end{eqnarray}

Precise knowledge of the intrinsic scatter in the above equations 
is a difficult task because it requires an accurate knowledge of the
measurement errors, which we do not claim to have.  As such, the scatter in
our current $M_{\rm bh}$--$L$ diagram probably cannot be used to help
distinguish between competing formation scenarios, as has been cleverly
proposed (e.g.\ Lahav et al.\ 2011; Shankar et al.\ 2012).  The measurement
errors should add in quadrature with the intrinsic scatter to give the total
rms scatter $\Delta$.  We can thus quickly proceed with a rough estimate of
the intrinsic scatter.  In terms of the vertical scatter about the $M_{\rm
bh}$--$\sigma$ relation, we have seen from Graham et al.\ (2011, their
Table~2) that $\epsilon \approx 3\Delta/4$.  Here we shall use a rough value
of 0.3 dex, which is consistent with the results in G\"ultekin et al.\
(2009b).
For the $M_{\rm bh}$--$L$ relation, Graham (2007a, his Table~4) reported a
similar value of around 0.3 dex for his predominantly luminous galaxy sample.
Obviously studies which have attempted to measure this (see also G\"ultekin et
al.\ 2009b) but failed to account 
for the bent nature of the $M_{\rm bh}$--$L$ relation will be in error at some level. 
We shall however use this value here for the core-S\'ersic galaxies. For the fainter
S\'ersic galaxies, which follow an $M_{\rm bh}$--$L$ relation that may be some
2.5 times steeper, we use a value of 0.75 dex.  This increase to the intrinsic
scatter in the vertical direction is expected if 
the intrinsic scatter in the horizontal direction, i.e.\ for the magnitudes, is equal for both
types of galaxy, i.e.\ core-S\'ersic and S\'ersic galaxies.

The above equations are used in Section~\ref{Sec_Int} where we predict a
number of black hole masses which are less than one million solar masses.

\subsection{Method of regression}

Park et al.\ (2012) explored how the {\sc BCES} linear regression code
from Akritas \& Bershady (1996) can produce a biased $M_{\rm bh}$--$\sigma$
relation if one's velocity dispersions have large measurement errors.  They
revealed that there is not a bias when these errors are less than $\sim$15\%, 
the uppermost value used in the literature. 
%
Park et al.\ (2012) additionally 
showed that the `forward' and `inverse' linear regression from the {\sc BCES}
code yield $M_{\rm bh}$--$\sigma$ relations which are consistent with those
derived using the `forward' and `inverse' versions of the modified {\sc
FITEXY} expression from Tremaine et al.\ (2002).  That there is agreement
between these codes in the $M_{\rm bh}$--$\sigma$ diagram is not a new result
(e.g.\ Novak et al.\ 2006; Graham \& Li 2009; Graham et al.\ 2011), but it may be a poorly
appreciated point worthy of some explanation before using the modified {\sc
FITEXY} code to check on our {\sc BCES}-derived $M_{\rm bh}$--$L$ relations. 

It is known that a linear regression which minimizes the offset of data from a
fitted line will typically yield a different result if one minimizes the
offset in either the vertical or the horizontal direction (e.g.\ Feigelson \&
Babu 1992).  It is also known that minimization of the vertical offsets (the
so-called `forward' regression) yields a shallower slope than obtained through 
minimization of the horizontal offsets (the `inverse' regression).  Obviously
neither of these are symmetrical regressions because there is a preferred
variable, that is, they do not treat the $x$ and $y$ data equally and
therefore they generate different fitted lines.  The two lines from each of
these non-symmetrical regressions can however be combined to produce a
symmetrical regression.  For example, the line which bisects the forward and
inverse regression lines, having the average angle between them, is the {\it
symmetrical} bisector regression line.
In general, symmetrical regressions are preferred when it is not known which
variable is dependent on the other, and when one wants to establish the
intrinsic relation between two quantities, the so-called theorist's question
(Novak et al.\ 2006).  Biases in the data sample, due to selection boundaries,
can of course bias the results from a symmetrical regression and thus create a
preference for either the forward or inverse regression depending on whether
the selection boundary is vertical or horizontal in one's diagram (e.g.\
Lynden Bell et al.\ 1988).
Typically in the literature, the modified {\sc FITEXY} routine has only been run as
a `forward' regression, minimizing the offsets in the vertical direction, where
as users of the {\sc BCES} code typically report the symmetrical bisector regression line.  It
is not appropriate to compare such results, and some confusion exists because
Tremaine et al.\ (2002) remarked that their modified {\sc FITEXY} routine 
treats the data symmetrically.  However if this was true, if the $x$ and
$y$ data was treated equally, then swapping the $x$ and $y$ data points would
not alter the fit that one obtains. 

Returning to our $M_{\rm bh}$--$L$ diagram, 
given the larger uncertainties on the spheroid magnitudes, compared to the
uncertainties on the velocity dispersions, we {\it have} used the 
modified {\sc FITEXY} routine to check if the {\sc BCES} bisector routine has yielded a
biased result.  In fitting the line $y = a + bx$, Tremaine et al.'s (2002)
modified version of the routine {\sc FITEXY} (Press et al.\ 1992, their
Section~15.3) minimizes the quantity
\begin{equation}
\chi^2 = \sum_{i=1}^N \frac{( y_i- a - bx_i)^2}
    { {\delta y_i}^2 + b^2{\delta x_i}^2 + \epsilon^2 }.
\label{Eq_One}
\end{equation}
The intrinsic scatter (in the $y$-direction) is denoted here by the term
$\epsilon$, and the measurement errors on the $N$ pairs of observables $x_i$
and $y_i$ are denoted by $\delta x_i$ and $\delta y_i$.  The intrinsic scatter
$\epsilon$ is solved for by repeating the fit with different values of
$\epsilon$ until $\chi^2/(N-2)$ equals 1. 
To achieve a minimization in the $x$-direction, i.e.\ to perform the `inverse'
regression, one can simply replace the $\epsilon^2$ term in the denominator 
with $b^2\epsilon^2$ (Novak et al.\ 2006; Graham 2007a; Graham \& Driver 2007a). 
The $M_{\rm bh}$--$L$ bisector line for the core-S\'ersic spheroids, obtained
from the bisector line of the forward and inverse modified {\sc FITEXY} routine,
has a slope of $-0.46$ and $-0.56$ in the $K$- and $B$- bands, respectively.
The slope of the $M_{\rm bh}$--$L$ bisector line for the S\'ersic spheroids,
obtained from the forward and inverse modified {\sc FITEXY} routine, is
$-0.95$ and $-0.88$ for the $K$ and $B$-bands, respectively.  These results
are consistent with the values obtained from the {\sc BCES} bisector routine,
as reported in Table~\ref{Tab3}.


\subsection{Further Pruning}

Curious readers may be wondering what happens if additional data points are
pruned from the $M_{\rm bh}$--$L$ diagram. 
The `core' galaxy with the faintest magnitude is the lenticular galaxy
NGC~3998.  With a velocity dispersion in excess of 300 km s$^{-1}$, 
this galaxy is expected to be a `core' galaxy.  Figure~8 from 
Gonzalez Delgado et al.\ (2008) reveals that this 
galaxy hosts an AGN that swamps the central light profile and any core which
may be there, hence the cautious `?'  on its classification in Table~1.
Re-labelling this galaxy to be a ``Sersic'' galaxy slightly steepens the
Sersic $M_{\rm bh}$--$L$ relation, and hence our case for a deviation from the previous
linear M-L relation, but not significantly. 
To the immediate lower left of this galaxy in the $M_{\rm bh}$--$L$ diagram is
NGC~4342.   This S\'ersic galaxy was labelled an outlier by Laor (2001) and
a potential outlier by H\"aring \& Rix (2004), both of whom excluded it from
their analysis.  Bogd\'an et al.\ (2012a) have also identified it as 
unusual, and it may be a heavily stripped galaxy (Blom et al.\ 2012), causing
it to deviate from the $M_{\rm bh}$--$L$ relation while still following the 
$M_{\rm bh}$--$\sigma$ relation.  M32 may be another such stripped galaxy
(Bekki 2001; Graham 2002). 
Excluding NGC~3998 and NGC~4342 results in an $M_{\rm 
bh}$--$L_{K_s}$ relation with slopes of $-0.50\pm0.08$ and 
$-1.07\pm0.24$ for the S\'ersic and core-S\'ersic galaxies, respectively.

\section{Discussion and Implications} 

\begin{figure}
\includegraphics[angle=0,scale=0.45]{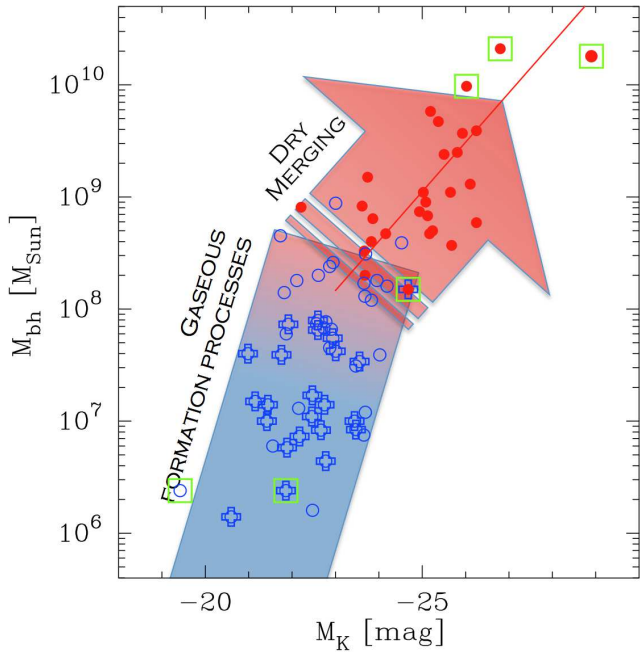}
\caption{
Schematic view of Figure~\ref{Fig2}b showing the evolutionary path of `dry'
galaxy mergers as they branch off from the steeper, near-quadratic, $M_{\rm
  bh}$--$L$ relation for S\'ersic galaxies built from `wet' mergers and/or 
gas-rich processes.  The upper red arrow 
roughly shows the direction of growth if $M_{\rm bh} \propto L^{1.0}$, while the slope
of the blue shaded area roughly follows $M_{\rm bh} \propto L^{2.5}$.
An additional five galaxies (NGC~3842 and NGC~4889 from McConnell et
al.\ 2011a, NGC~1316, NGC~3079 and M32), which were 
excluded from the regression analysis in Figure~\ref{Fig2}, 
are shown here, along with OJ278 from Valtonen et al.\ (2012), 
and marked with a green square. 
}
\label{Fig3}
\end{figure}

The coevolution of black holes and their host spheroids has resulted in a
bent, i.e.\ a non-(log-linear), $M_{\rm bh}$--$L$ relation which can be
approximated by a broken power-law having an exponent of $\sim$1
for spheroids with core-S\'ersic profiles and $\sim$2.5 for spheroids with
S\'ersic profiles. 
%
%
The major, i.e.\ near equal mass, merger of gas-free galaxies near the
bright-end of the S\'ersic sequence will preserve the $M_{\rm bh}/L$ ratio and
create the core-S\'ersic galaxies, which are observed to follow a near linear
$M_{\rm bh}$--$L$ relation.  That is, such dry mergers do not move galaxies
along the steep $M_{\rm bh}$--$L$ relation defined by S\'ersic galaxies, but
drive them off it to form a distinctly different sequence
(Figure~\ref{Fig3})\footnote{A substantial gas-rich merger with, or gas inflow
onto, a core-S\'ersic galaxy may drive it above the one-to-one relation seen
in the $M_{\rm bh}$--$L$ diagram.}. 
%
The above picture (and Figure~\ref{Fig3}) suggests that the bulk of AGN
emission, i.e.\ black hole growth through gas-rich processes, should be
associated with black holes having masses less than a few times $10^8
M_{\odot}$.  The peak of the Eddington ratio at $M_{\rm bh}=$(4--8)$\times10^7
M_{\odot}$ is not inconsistent with this (DeGraf et al.\ 2012).
Magnitude-limited high-redshift quasar surveys may thus only be probing the tip of
the proverbial iceberg. 

An exponent of 1 for the core-S\'ersic galaxies in the $M_{\rm bh}$--$L$ 
diagram would mean that there is a constant $M_{\rm bh}/M_{\rm spheroid}$ mass ratio for
these galaxies.
%
For a core-S\'ersic spheroid with a (dust-free) absolute magnitude
$M_{K_s} = -25.33$ mag, and assuming an $M/L_{K}$ ratio of 0.8 (Bell \& de
Jong 2001) and a solar $K$-band absolute magnitude of 3.27 mag (Cox 2000), we
find a median mass ratio $M_{\rm bh}/M_{\rm Sph,*} = 0.49$\%.
For a core-S\'ersic spheroid with a (dust-free) absolute magnitude $M_B
= -21$ mag, and assuming an $M/L_B$ ratio of 5 and a solar $B$-band absolute
magnitude of 5.47 mag (Cox 2000), we find a similar (black hole)-to-(stellar)
mass ratio of 0.55\%.  
These ratios are larger than the old median value of 
0.14--0.2\% (e.g.\ Ho 1999; Marconi \& Hunt 2003; H\"aring \& Rix 2004) for
the (black hole)-to-(spheroid dynamical) mass ratio.  
Even if we assumed $M/L_B = 8$, we find $M_{\rm bh}/M_{\rm Sph,*} = 0.35$\%.

The steep exponent of $\sim$2.5 for the S\'ersic galaxies in the $M_{\rm
bh}$--$L$ diagram means that the pre-existing single log-linear $M_{\rm
bh}$--$L$ relation will overestimate the black hole masses in galaxies whose
black hole mass is less than $\sim$$10^8 M_{\odot}$. This is in fact already
evident in Khorunzhev et al.\ (2012) who 
found that $M_{\rm bh}$ was overestimated using the old $M_{\rm bh}$--$L$
relations for $M_{\rm bh} < 10^8 M_{\odot}$, just where the bend in the 
$M_{\rm bh}$--$L$ relation sets in.
Similarly, Mathur et al.\ (2012, their figure~3) 
show ten galaxies below the old $M_{\rm bh}$--$L$ relation at $\log(M_{\rm
  bh}) = 7\pm0.4$. 
However their interpretation that these galaxies host pseudobulges, 
based on their location in the $M_{\rm bh}$--$L$ diagram, is premature
because these galaxies are found where the classical bulges and elliptical
galaxies reside. 

Having the correct black hole scaling relations is, obviously, important when
trying to predict a galaxy's black hole mass.  Equally important is having the
correct velocity dispersion or bulge magnitude. 
In regard to the core-S\'ersic galaxy NGC~4382 (Dullo \& Graham 2012),
G\"ultekin et al.\ (2011) predicted a black hole mass that differed by an
order of magnitude when one used the $M_{\rm bh}$--$L$ and $M_{\rm
bh}$--$\sigma$ relations (G\"ultekin et al.\ 2009b).
Here we predict $\log(M_{\rm bh}/M_{\odot}) = 8.0\pm0.3$ when using $\sigma =
182\pm5$ km s$^{-1}$ (G\"ultekin et al.\ 2011), in agreement with G\"ultekin
et al.\ (2011).  However this galaxy appears to be a disturbed, face-on,
non-barred, lenticular galaxy (Laurikainen et al.\ 2011) and we therefore do
not use the total galaxy light to estimate its central black hole mass.
The total 2MASS $K_s$-band magnitude of 6.145 mag corresponds to an absolute
magnitude of roughly $-25.1$ mag given a distance modulus of 31.27 (Tonry et
al.\ 2001; Blakeslee et al.\ 2002).  Galactic extinction and redshift dimming
are negligible.  Using an average, dust-corrected, S0 galaxy bulge-to-disk
flux ratio of one third, we thus have a $K_s$-band bulge magnitude of $-23.9$
mag.  Using the $M_{\rm bh}$--$L_{K_s}$ relation
in Table~\ref{Tab3}, we find $\log(M_{\rm bh}/M_{\odot}) = 8.6\pm0.5$ when using
a factor of 2 uncertainty on this bulge magnitude.  This estimated black hole
mass has over-lapping error bars with the first estimate.  We would however predict a
black hole mass of $\sim 10^9 M_{\odot}$, in agreement with G\"ultekin et al.\
(2011), if all of the light in this disturbed galaxy was assigned to a single
spheroid component.
%
%
%

Having the right $M_{\rm bh}$--$L$ relation is of course also important when
using the galaxy/spheroid luminosity function to predict the black hole mass
function, and subsequently the black hole mass density (e.g.\ Graham \& Driver
2007b, and references therein).  Knowledge of the correct $M_{\rm bh}$--$L$
relation will additionally be useful for cosmological / evolutionary studies
of the nucleus-to-(host spheroid) mass ratio (e.g.\ Kisaka \& Kojima 2010;
Schulze \& Wisotzki 2011; Portinari et al.\ 2012; Sesana 2012), and for many
other studies.
The consequences of the new bent $M_{\rm bh}$--$L$ and bent $M_{\rm
bh}$--$M_{\rm spheroid}$ relations are widespread and dramatic.  Below we discuss
some of these in more detail.

\subsection{Intermediate mass black holes}\label{Sec_Int}

Classical elliptical galaxies with $M_B \approx -18$ mag will now have
($M_{\rm bh}$--$L$)-derived black hole masses which are ten times lower than
previously predicted.
%
Similarly, we also now have revised predictions for the location of $10^6
M_{\odot}$ black holes within the $M_{\rm bh}$--$L$
diagram.  Based on their host spheroid luminosity, they will have black hole
masses which are roughly 10 times lower than predicted from the extrapolation
of the $M_{\rm bh}$--$L$ relation defined by luminous core-S\'ersic galaxies,
placing them near the current detection limit shown in Figure~\ref{Fig2}.
Black holes with masses of $10^5$--$10^6 M_{\odot}$ will be even further from
the extrapolated core-S\'ersic $M_{\rm bh}$--$L$ relation.  A number
of AGN residing in elliptical galaxies, not to be confused with the bulges of
disk galaxies, have been observed below this relation.  For example, Greene et
al.\ (2008) showed some AGN which had black hole masses that are an order of
magnitude below both the old $M_{\rm bh}$--$L$ relation and the $M_{\rm
bh}$--$M_{\rm spheroid}$ relation from H\"aring \& Rix (2004).  Our
interpretation of this offset differs from Greene et al.\ (2008), in that we consider
their sample of non-disk galaxies to be the extension of the S\'ersic
sequence, rather than a separate galaxy population which 
Greene et al.\ (2008) refer to as ``blobs''.

\subsubsection{Individual galaxies}

Interestingly, the ($M_{\rm bh}$--$M_{\rm spheroid}$)-derived mass determination
(using the expression from H\"aring \& Rix 2004) for the black hole associated
with the recent ultraviolet-optical flare from the tidal disruption of a
helium-rich stellar core in a galaxy at a redshift of 0.17 (Gezari et al.\
2012) can now be revised.  Using the updated $M_{\rm bh}$--$M_{\rm spheroid}$
relation from Graham (2012a), its black hole mass is $\sim$5 times lower than
the reported value of $4\times10^6 M_{\odot}$, making it slightly less than
one million solar masses. 

Our revised black hole scaling relations suggest that the high-energy 
transient Sw 1644+57 (Levan et al.\ 2011; Bloom et al.\ 2011) may have been
associated with a stellar capture event by a black hole having a mass of 
$0.6\times10^6 M_{\odot}$ (and even less if the host galaxy is a disk galaxy). 
Not only does the host galaxy's absolute magnitude ($M_V = -18.19$ mag)
suggest this low black hole mass (for $B-V = 0.9$), but so does the reported
host galaxy mass of a few billion solar masses (Bloom et al.\ 2011) when using
the $M_{\rm bh}$--$M_{\rm spheroid}$ relation from Graham (2012a).  We
therefore tentatively claim the presence of an intermediate mass black hole in
this galaxy, adding to the detection of other intermediate mass black hole
candidates (e.g., Farrell et al.\ 2009; Soria et al.\ 2010; Sutton et al.\
2012).

The revised $M_{\rm bh}$--$L$ scaling relations presented here give rise to
the need to re-derive many past estimates of black hole masses, in particular
those associated with active galactic nuclei in low luminosity bulges.  
Whether or not these objects can provide unique observational constraints on
the progenitors of {\it supermassive} black holes is unclear, but they do
narrow the divide between today's supermassive black holes and their `seeds'.

For the Seyfert Sc galaxy NGC~3367 (McAlpine et al.\ 2011), Dong \& De
Robertis (2006) report a $K$-band bulge magnitude of $-21.6$ mag,
corresponding to a bulge-to-total flux ratio of 0.08.  Hern\'andez-Toledo et
al.\ (2011) find that the bulge has a S\'ersic index between 1 and 1.8, and
after including a bar in their image analysis they report a bulge-to-disc
ratio of 0.07 to 0.11.
The bulge magnitude from Dong \& De Robertis (2006), after brightening it by
0.11 mag for dust (Driver et al.\ 2008), yields an expected black hole mass of
$\log M_{\rm bh}/M_{\odot} = 6.2\pm0.9$.
McAlpine et al.\ (2011) also studied the Sbc galaxy NGC~4536, reported to have
a bulge S\'ersic index of $1.88\pm0.35$ by Fisher \& Drory (2008), and an 
absolute $V$-band magnitude of -18.4 mag assuming a distance of 25.3 Mpc. 
With a disk inclination of 62 degrees (Fisher \& Drory 2008) we apply a dust
correction of 1.3 mag (Driver et al.\ 2008), and using an assumed $B-V$ bulge color of
0.8 we have that $M_B = -18.9$ mag.  The expected black hole mass is thus
$\log M_{\rm bh}/M_{\odot} = 7.3\pm0.8$, assuming a 0.3 mag uncertainty on
the bulge magnitude.  These two galaxies are not, therefore, expected to house low
intermediate mass black holes. 

If the dwarf galaxy Pox~52 --- long thought to house an AGN (Kunth, Sargent \&
Bothun 1987) --- is an elliptical galaxy, then its $B$-band magnitude of
$-16.8$ mag (Barth et al.\ 2004) translates into an expected black hole mass
of $\log M_{\rm bh}/M_{\odot} = 5.3\pm0.9$, with the uncertainty
of 0.5 dex derived assuming an uncertainty of 0.3 mag on the galaxy
magnitude. 

For the nearby dwarf lenticular galaxy NGC~404, the Galactic extinction
corrected $H$-band magnitude of $-$19.65 mag from Seth et al.\ (2010) becomes
$M_{K_s} = -19.85$ mag when using an $H-K_s$ color of 0.2 mag (e.g.\ Balcells
et al.\ 2007).  Thought to house an AGN (e.g.\ Nyland et al.\ 2012, and
references therein), the expected black hole mass in NGC~404 is $\log M_{\rm
  bh}/M_{\odot} = 4.1\pm1.1$, using a magnitude error of 0.3 mag. 
For comparison, although restricted by assumptions about the inner stellar
mass-to-light ratio, the stellar dynamical modeling from Seth et al.\ (2010) 
suggests that NGC~404 has a black hole mass less than $10^5 M_{\odot}$. 

The bulge of the Scd galaxy NGC~1042 has a dust corrected magnitude of
$-19.13\pm0.75$ $K$-mag (Knapen et al.\ 2003; Graham \& Worley 2008) and
houses an active black hole (Shields et al.\ 2008; Seth et al.\ 2008).  From
the new $K$-band $M_{\rm BH}$-$L_{\rm bulge}$ relation, the expected black
hole mass is $\log M_{\rm bh}/M_{\odot} = 3.3\pm1.4$.
 
Given a $B$-band bulge magnitude of $-13.93$ mag (Graham \& Spitler 2009) 
for the AGN-hosting galaxy NGC~4395 (Filippenko \& Ho 2003), 
the expected black hole mass in this late-type galaxy's bulge
is equally small. Assuming a 0.75 mag uncertainty on the bulge magnitude, 
the black hole mass is $\log M_{\rm bh}/M_{\odot} = 2.6\pm1.3$. 
Allowing for the 1-sigma uncertainty given here, this mass resides just below
the $10^4$--$10^5 M_{\odot}$ mass range estimated by Filippenko \& Ho (2003) and 
is also below the (reverberation mapping)-derived estimate of 
$3.6\pm2.2 \times 10^5 M_{\odot}$ 
(Peterson et al.\ 2005), even after it has been halved to $1.8\pm0.6 \times 10^5 M_{\odot}$ 
due to the updated virial $f$-factor in Graham et al.\ (2011).  
A lower black hole mass is however expected from the short
time-scale variations of the X-ray flux which can vary by an 
order of magnitude (Vaughan et al.\ 2005), 
and would of course necessitate a higher Eddington accretion rate than the 
previously assumed maximum of just a couple of percent. 
%

All of our black hole mass estimates above have been collated in
Table~\ref{Tab5} for ease of reference and overview.

\begin{table}
\begin{center}
\caption{Supermassive and intermediate mass black holes}
\label{Tab5}
\begin{tabular}{@{}lll@{}}
\hline
Ident.         &  magnitude       &  $\log M_{\rm bh}$ \\
\hline  
NGC~4536       &  $-18.9\, B$-mag   &  $7.3\pm0.8$ \\ 
NGC~3367       &  $-21.71\, K$-mag  &  $6.2\pm0.9$ \\
Gezari et al.\ &  ...               &  $5.9\pm0.9$ \\
Sw 1644+57$^*$ &  $-18.19\, V$-mag  &  $5.8\pm0.9$ \\
Pox~52         &  $-16.8 \, B$-mag  &  $5.3\pm0.9$ \\ 
NGC~404        &  $-19.85\, K$-mag  &  $4.1\pm1.1$ \\
NGC~1042       &  $-19.13\, K$-mag  &  $3.3\pm1.4$ \\
NGC~4395       &  $-13.93\, B$-mag  &  $2.6\pm1.3$ \\
\hline
\end{tabular}

$^*$ This transient's designation is used to identify the host galaxy.
\end{center}
\end{table}


Typically, when X-ray flux is detected in galaxies having a candidate
intermediate mass black hole, the level of emission is so low that skeptics
point to possible stellar origins for its production.  That is, in general,
there is not the danger that such galaxies might have too much X-ray
luminosity supportive of a supermassive black hole rather than an 
intermediate mass black hole.  
%
%
%

\subsubsection{A catalog of galaxies: Dong \& De Robertis}

In addition to the above mentioned galaxies which have had individual papers
dedicated to them, we are able to identify many other intermediate mass black
hole candidates from a catalog of low-luminosity AGN.  The log-linear
$K_s$-band $M_{\rm bh}$--$L$ relation of Dong \& De Robertis (2006), which was
consistent with that from Marconi \& Hunt (2003), was used by Dong \& De
Robertis to predict the masses of black holes in 117 disc galaxies hosting
low-luminosity AGN.  From their sample of bulge luminosities, they reported
only 4 galaxies with estimated black hole masses less than $10^6 M_{\odot}$
and such that $\log M_{\rm bh}/M_{\odot}$ = 5.6 to 5.8.  Given this, coupled
with the uncertainty in the $M_{\rm bh}$--$L$ relation, they understandably
paid no attention to the possible detection of intermediate mass black holes
in their sample.  However given the dramatically steeper $M_{\rm bh}$--$L$
relation reported in this paper for intermediate- and low-luminosity
spheroids, we expect that many of the black hole mass estimates in Dong \& De
Robertis will have been overestimated.  As a consequence, we note in passing
that the associated scaling relations reported there between black hole mass
and line width (FWHM [N {\sc II}]), as well as emission-line ratios (e.g., [O
{\sc III}]/H$\beta$, [O {\sc I}]/H$\alpha$, [N {\sc II}]/H$\alpha$, and [S
{\sc II}]/H$\alpha$), may need to be re-derived.

Here we predict new black hole masses for the 41 galaxies reported by Dong \&
De Robertis (2006, their Table~4) to have absolute $K_s$-band bulge magnitudes
fainter than $-21.0$ mag.  Before applying our S\'ersic $M_{\rm bh}$--$L$
relation, we correct (brighten) these bulge magnitudes for internal dust
extinction.  This has been done using the inclination-dependent, dust
correction from Driver et al.\ (2008), and is shown in Table~\ref{Tab6}.  The
predicted black hole masses shown there {\it assumes} that the $M_{\rm
bh}$--$L$ relation can be extrapolated to lower black hole masses, and the
quoted error was derived assuming an uncertainty on the magnitudes of 0.3 mag.

The two lowest mass, intermediate mass black hole candidates in
Table~\ref{Tab6} are the barred galaxies NGC~4136 ($\log M_{\rm bh}/M_{\odot}
= 2.2 \pm 1.3$) and NGC~3756 ($\log M_{\rm bh}/M_{\odot} = 2.6 \pm 1.3$).  We
can apply our updated $M_{\rm bh}$--$\sigma$ relation for barred galaxies,
shown in bold in Table~\ref{Tab3}, although a serious concern is that the
available velocity dispersions of 38.4$\pm$8.7 and 47.6$\pm$8.5 (Ho et al.\
2009) may be elevated by the velocity dispersion of the disk given the low
bulge luminosities.  The respective ($M_{\rm bh}$--$\sigma$)-derived masses
$\log M_{\rm bh}/M_{\odot} = 4.13 \pm 1.26$ and $ 4.62 \pm 1.11$ do however
still have over-lapping error bars with the ($M_{\rm bh}$--$L$)-derived
masses.

Finally, we note that moving down the spheroid luminosity function, black holes
appear to give way to, or at least co-exist with, dense nuclear star clusters
(e.g.\ Graham \& Spitler 2008; Graham 2012b; Neumayer \& Walcher 2012; Leigh,
B\"oker \& Knigge 2012; Scott \& Graham 2012).  It is not clear to what lower
masses our S\'ersic $M_{\rm bh}$--$L$ relation may hold.   We do however note
that many new intermediate mass black hole candidates, with $M_{\rm bh} = 
0.8\times10^5$--$10^6 M_{\odot}$, measured from virial mass estimator based on
the broad H$\alpha$ line, have recently been identified (Dong et al.\ 2012). 

\begin{table}
\begin{center}
\caption{Predicted intermediate mass black holes}
\label{Tab6}
\begin{tabular}{@{}lccl@{}}
\hline
Galaxy  &  $M_{K_s,orig}$  &  $M_{K_s,corr}$  &  $\log M_{\rm bh}/M_{\odot}$ \\
\hline
NGC 3185  & -20.80 & -20.94 & 5.3$\pm$0.9 \\
NGC 3593  & -20.70 & -21.03 & 5.4$\pm$0.9 \\
NGC 3600  & -19.20 & -19.72 & 4.0$\pm$1.1 \\
NGC 3729  & -20.10 & -20.24 & 4.6$\pm$1.0 \\
NGC 4245  & -20.80 & -20.93 & 5.3$\pm$0.9 \\
NGC 4314  & -21.00 & -21.11 & 5.5$\pm$0.9 \\
NGC 4369  & -20.90 & -21.01 & 5.4$\pm$0.9 \\
NGC 4470  & -20.40 & -20.53 & 4.9$\pm$1.0 \\
NGC 3003  & -19.60 & -20.09 & 4.4$\pm$1.0 \\
NGC 3043  & -20.80 & -21.17 & 5.6$\pm$0.9 \\
NGC 3162  & -20.00 & -20.12 & 4.4$\pm$1.0 \\
NGC 3344  & -19.30 & -19.41 & 3.7$\pm$1.1 \\
NGC 3507  & -20.90 & -21.01 & 5.4$\pm$0.9 \\
NGC 3684  & -20.60 & -20.74 & 5.1$\pm$1.0 \\
NGC 3686  & -20.60 & -20.72 & 5.1$\pm$1.0 \\
NGC 3756  & -18.20 & -18.43 & 2.6$\pm$1.3 \\
IC 467    & -19.20 & -19.50 & 3.8$\pm$1.1 \\
NGC 514   & -19.90 & -20.02 & 4.3$\pm$1.0 \\
NGC 628   & -20.40 & -20.51 & 4.9$\pm$1.0 \\
NGC 864   & -19.90 & -20.03 & 4.3$\pm$1.0 \\
NGC 2276  & -20.90 & -21.01 & 5.4$\pm$0.9 \\
NGC 2715  & -19.20 & -19.56 & 3.8$\pm$1.1 \\
NGC 2770  & -19.10 & -19.50 & 3.8$\pm$1.1 \\
NGC 2776  & -20.90 & -21.01 & 5.4$\pm$0.9 \\
NGC 2967  & -20.30 & -20.41 & 4.8$\pm$1.0 \\
NGC 3041  & -19.90 & -20.05 & 4.4$\pm$1.0 \\
NGC 3198  & -19.80 & -20.11 & 4.4$\pm$1.0 \\
NGC 3359  & -20.80 & -20.97 & 5.4$\pm$0.9 \\
NGC 3430  & -20.10 & -20.29 & 4.6$\pm$1.0 \\
NGC 3433  & -20.40 & -20.51 & 4.9$\pm$1.0 \\
NGC 3486  & -19.90 & -20.03 & 4.3$\pm$1.0 \\
NGC 3596  & -21.00 & -21.11 & 5.5$\pm$0.9 \\
NGC 3666  & -18.20 & -18.63 & 2.8$\pm$1.2 \\
NGC 3726  & -20.10 & -20.24 & 4.6$\pm$1.0 \\
NGC 3780  & -20.20 & -20.32 & 4.7$\pm$1.0 \\
NGC 3938  & -20.50 & -20.61 & 5.0$\pm$1.0 \\
NGC 4062  & -18.50 & -18.78 & 3.0$\pm$1.2 \\
NGC 4096  & -19.40 & -19.84 & 4.1$\pm$1.1 \\
NGC 4136  & -18.00 & -18.11 & 2.2$\pm$1.3 \\
NGC 4152  & -20.80 & -20.92 & 5.3$\pm$0.9 \\
NGC 4212  & -20.10 & -20.27 & 4.6$\pm$1.0 \\
\hline
\end{tabular}

$M_{K_s,orig}$ is each galaxy's $K_s$-band, spheroid magnitude reported by Dong \& De
Robertis (2006). $M_{K_s,corr}$ is the dust-corrected magnitude following
Driver et al.\ (2008).  $M_{\rm bh}$ is our predicted black hole mass using
the S\'ersic $M_{\rm bh}$--$L_{K_s}$ relation from Table~\ref{Tab4}. 
(For reference, Dong \& De Robertis (2006) predicted only 4 galaxies to have 
black hole masses less than $10^6 M_{\odot}$, which were in the range $\log
M_{\rm bh}/M_{\odot} = 5.6$ to 5.8.)

\end{center}
\end{table}

\subsection{The $M_{\rm bh}$--$M_{\rm host-galaxy}$ connection}

For early-type galaxies, the dynamical mass-to-light ratio is thought to roughly change with the
optical spheroid luminosity as $(M/L)_{\rm dynamical} \propto L^{1/4}$ (e.g.\
Faber et al.\ 1987; Cappellari et al.\ 2006, their eq.~17) or rather 
$\propto L^{1/3}$ (Cappellari et al.\ 2006, their eq.9 and 11), while $M/L$ should
obviously be constant for the core-S\'ersic galaxies if built via dry
(additive) mergers.  
That is, the core-S\'ersic $M_{\rm bh}$--$M_{\rm Sph,dyn}$ relation should have the same
(close to unity) slope as the core-S\'ersic $M_{\rm bh}$--$L_{\rm spheroid}$
relations 
$M_{\rm bh} \propto L_{B}^{1.35\pm0.30}$ and $M_{\rm bh} \propto L_{K_s}^{1.10\pm0.20}$.
However we need to caution that galaxy samples with a greater fraction 
of core-S\'ersic galaxies relative to S\'ersic galaxies may have in the past resulted in a
shallower {\it average} $(M/L)_{\rm dynamical} \propto L$ relation than is
actually applicable to the S\'ersic galaxies on their own.
Studies of the Fundamental Plane (Djorgovski \& Davis 1987) in the
near-infrared $J$- and $K$-bands, which are less affected by stellar
population differences across the galaxy luminosity range, have tended to
yield $(M/L)_{\rm dynamical} \propto L_K^{1/6}$ (e.g.\ Magoulas et al.\ 2012;
La Barbera et al.\ 2010), although the expanded SAURON survey with 72 E/S0/Sa
galaxies reported $(M/L)_{\rm dynamical} \propto L_V^{1/3}$ and $\propto
L_{3.6\mu}^{1/4}$ (Falc\'on-Barroso et al.\ 2011).
The ATLAS$^{\rm 3D}$ survey (Cappellari et al.\ 2011) is comprised of just a
few percent of core-S\'ersic galaxies supported by random motions rather than
rotational velocity (Emsellem et al.\ 2011) and it will therefore be
interesting to see if they also find these steeper $M/L$--$L$ relations.

For now, using the shallower $M/L$ relations above, which admittedly may
underestimate the slope of the S\'ersic galaxy $(M/L)_{\rm dynamical} \propto
L$ relation, our S\'ersic $M_{\rm bh}$--$L_{\rm spheroid}$ expressions can be roughly
transposed into (black hole)--(host spheroid) mass relations.  
One has that 
$M_{\rm bh} \propto L_{B}^{2.35\pm0.40}$ maps into 
$M_{\rm bh} \propto M^{1.88\pm0.32}_{\rm Sph,dyn}$ (when using $M/L_B \propto L_B^{1/4}$) and 
$M_{\rm bh} \propto L_{K_s}^{2.73\pm0.55}$ maps into 
$M_{\rm bh} \propto M^{2.34\pm0.47}_{\rm Sph,dyn}$ (when using $M/L_{K_s} \propto L_{K_s}^{1/6}$). 
This can be directly compared with the slopes of 1.01$\pm$0.52 for the
core-S\'ersic galaxies in Graham (2012a), and 2.30$\pm$0.47 and 1.92$\pm$0.38 
for the S\'ersic and non-barred S\'ersic galaxies in that study. 

Rather than described by a single power-law (e.g.\ Magorrian et al.\ 1998),
the full $M_{\rm bh}$--$M_{\rm Sph,dyn}$ relation appears to be better
described by a broken power-law having exponents of $\sim$1 for the
core-S\'ersic spheroids and $\sim$2 for the S\'ersic spheroids, in agreement
with the previous announcement by Graham (2012a) based on a re-analysis of the
H\"aring \& Rix (2004) data (see also Balcells et al.\ 2007, their
equation~13).
The non-linear, albeit still a single log-linear, (black hole)-(host stellar
mass) relation from Laor (2001) was given as $M_{\rm bh} \propto M_{\rm
Sph,*}^{1.53\pm0.14}$ and was not a mistake but rather arose from fitting a
sample of what are undoubtedly S\'ersic and core-S\'ersic spheroids.

If, as commonly believed, the core-S\'ersic galaxies are built from the simple
addition of high-mass gas-free S\'ersic galaxies and/or other core-S\'ersic
galaxies, then one may similarly expect $M_{\rm bh}$ to be linearly correlated
with the dark halo mass in core-S\'ersic galaxies (Ferrarese 2002; Bogd\'an et
al.\ 2012a).  Of course this (non-causal) relation will not hold across the
full mass range of S\'ersic spheroids (Ho 2007).
%
The single power-law relation involving the mass of the central 
black hole and the surrounding dark matter halo which was derived for a sample
of both S\'ersic and 
core-S\'ersic galaxy types, $M_{\rm bh} \propto M_{\rm
DM}^{1.65 \, - \, 1.82}$ (Ferrarese 2002), may be better described by a
broken power-law --- at least over the high-mass range where a correlation exists. 
Looking at Figure~5 from Ferrarese (2002), galaxies 
with $M_{\rm bh} \gtrsim 10^8 M_{\odot}$ appear to be better described by a relation
with an exponent closer to 1 than 1.65--1.82, with a quick analysis of their
Figure revealing that an exponent of $\approx1.2$ provides a fair fit.  This
is close to the values of 1.10 and 1.35 from our relations $M_{\rm bh} 
\propto L_{K_s}^{1.10\pm0.20}$ and $M_{\rm bh} \propto L_{B}^{1.35\pm0.30}$.
%
Thankfully complex models involving the growth of seed black holes in high-redshift dark
matter halos 
are not required to explain this near-linear relation if it is
created by the simple addition of similar galaxies.

\subsection{Black hole growth in semi-analytical models}

Given the log-linear relation between S\'ersic index $n$ and spheroid
luminosity $L$ (e.g.\ Graham \& Guzm\'an 2003, and references therein), the
bent $M_{\rm bh}$--$n$ relation (Graham \& Driver 2007a)
necessitates that the $M_{\rm bh}$--$L$ relation must, as observed here, also
be bent.
It would appear that black holes within spheroids formed from wet-mergers and
merger triggered accretion episodes (Sanders \& Mirabel 1996; 
Villar-Mart\'{\i}n et al.\ 2012), and/or monolithic-like collapse ---
especially for isolated systems --- grew via quadratic black hole growth such that
the $M_{\rm bh}$--$M_{\rm spheroid}$ relation has a power-law slope of 2 for the
S\'ersic spheroids.  That is, the black hole should grow more quickly than the
population of stars in the host spheroid, and this has in fact recently been
observed (Seymour et al.\ 2012; see also Greene et al.\ 2009).  

Following Kauffmann \& Haehnelt (2000), 
numerous theories and models (e.g.\ Hopkins et al.\ 2006; Johansson, Naab \&
Burkert 2009; Cen 2012; Hirschmann et al.\ 2012) 
have been developed to create or match 
a near-linear $M_{\rm bh}$--$L$ and $M_{\rm bh}$--$M_{\rm spheroid}$ relation. 
Unfortunately because of this, some of their 
claimed `success' implies that they are actually in need of refinement.  
However some models may contain important elements or clues that have 
been somewhat over-looked by the community. 
For example, the models of Dubois et al.\ (2012) reveal
that while the $M_{\rm bh}$--$\sigma$ relation is linear and has a steeper
slope than 4, their simulated $M_{\rm bh}$--$L$ relation is bent at around
$M_{\rm bh} = 10^8 M_{\odot}$, 
as is the simulated $M_{\rm bh}$--$M_{\rm Spheroid}$ relation of Cirasuolo et
al.\ (2005)\footnote{Cirasuolo et al.\ (2005) additionally predict a steepening of the
  $M_{\rm bh}$--$\sigma$ relation at lower black hole masses.}. 
The hydrodynamical simulation models of Khandai et al.\ (2012, see their 
figure~7) clearly reveal a steep $M_{\rm bh}$--$M_{\rm spheroid}$ relation such that
black hole growth outpaces that of the host spheroid, in agreement with Graham
(2012a) and this study. 

The primary, so-called `quasar' or `cold', mode of black hole growth during
gas-rich mergers / processes currently assumes, in semi-analytical models (Kauffmann \&
Haehnelt 2000, their eq~2; Croton et al.\ 2006, their eq.~8; Guo et al.\ 2011,
their eq.~36)\footnote{For clarity we note that Guo et al.\ (2011) excluded
the square on the normalized velocity term in their eq.~36.}, 
that black hole growth occurs via accretion which is linearly proportional to the
inflowing mass of cold gas (which also produces the host spheroid), modulated 
by an efficiency which is lower for both unequal 
mass mergers (Croton et al.\ 2006) and less massive (more gas-rich) 
systems with lower virial velocities.  We therefore offer the following new
prescription for the increase in black hole mass, $\delta M_{\rm bh}$, due to
gas accretion during wet mergers: 
%
%
%
\begin{equation}
\delta M_{\rm bh} \propto 
\left(\frac{M_{\rm min}}{M_{\rm maj}}\right)
\left[ \frac{M^2_{\rm cold}}{1+(280\, {\rm km\, s}^{-1})/V_{\rm
      virial}}\right]. 
\label{EqQ}
\end{equation} 
Here, $M_{\rm min}$ and $M_{\rm maj}$ are the total baryonic masses from the
minor and major galaxies involved, and $M_{\rm cold}$ is their combined cold
gas mass. The term $V_{\rm virial}$ is the circular or `virial' velocity of the
newly wed galaxy's halo, normalized at 280 km s$^{-1}$ following Kauffmann \&
Haehnelt (2000).
This expression, now with an exponent of 2 on the cold gas mass, 
assumes quadratic growth for the black hole relative to the
stellar mass of the host spheroid. 

Another consequence of this order-of-magnitude lower $M_{\rm bh}/M_{\rm spheroid}$
mass ratio in faint spheroids is that AGN feedback, and in particular the
ratio of this feedback energy ($\propto M_{\rm bh}$) to the binding energy of
the spheroid ($\propto M_{\rm spheroid}\sigma^2$), will be an order of magnitude
less than previously assumed by studies which used the old constant value of
$M_{\rm bh}/M_{\rm spheroid}\approx 0.002$ (Marconi \& Hunt 2003; H\"aring \& Rix
2004).

\subsection{Barred galaxies}


As noted earlier, barred galaxies can reside up to $\sim$1 dex below, or
perhaps rather to the right of, 
non-barred galaxies in the $M_{\rm bh}$--$\sigma$ diagram, with the mean
vertical offset 0.3 to 0.4 dex (e.g.\ Graham 2008a; Hu 2008).
Galaxies with AGN have also been shown to display this same behavior 
in the $M_{\rm bh}$--$\sigma$ diagram 
(e.g.\ Graham \& Li 2009, see also Wandel 2002; Mathur \& Grupe 2005; 
Wu 2009; Mathur et al.\ 2012, while alternate views can be found in 
Botte et al.\ 2005; Komossa \& Xu 2007; Decarli et al.\ 2008a and Marconi et al.\ 2008). 
This is not to say that all barred galaxies are offset from the {\it barless} 
$M_{\rm bh}$--$\sigma$ relation, only that a significant fraction are.  In the analysis by Hu
(2008), he considered the offset galaxies to be 'pseudobulges', although 
all of his offset galaxies were also barred galaxies.  Greene, Ho, \& Barth
(2008, see their figure~7b) subsequently showed that their sample of 11 
`pseudobulges' was {\it not} offset from the $M_{\rm bh}$--$M_{\rm spheroid}$ relation
defined by massive classical systems\footnote{We speculate that this lack of
  an offset may 
have been due to an underestimation of dust obscuration from the disks
surrounding their bulges.}. (Oddly, their work is often mis-quoted as
evidence that pseudobulges in disk galaxies are an offset population.) 
With new and updated data, we have observed that 
the mean offset of (in this case 21) barred galaxies in the $M_{\rm bh}$--$\sigma$ diagram
remains.  These galaxies also appear to follow a relation which is parallel
with that defined by luminous core-S\'ersic galaxies.  The rather
narrow baseline, of effectively only $\sim$2 mag, for the 11 `pseudobulges' in
Kormendy et al.'s (2011) $M_{\rm bh}$--$L$ diagram 
may partly (see also Graham 2011b) explain why they did not observe a 
correlation between black hole mass and bulge luminosity. 
It is not yet clear if barred galaxies are offset from non-barred 
(S\'ersic) galaxies in the $M_{\rm bh}$--$M_{\rm spheroid}$ diagram. 


Graham (2008a) had previously noted the connection between
bar instabilities and pseudobulges\footnote{Whether or not the bulges in these
galaxies formed solely from a redistribution of disk stars, to produce
`pseudobulges', is unknown due to the many difficulties in identifying
pseudobulges (Graham 2011a, 2012c).}, and wrote that if the offset barred
galaxies have discrepantly low black hole masses rather than elevated $\sigma$-values,
then they should also appear as systematic outliers in the $M_{\rm bh}$--$L$
diagram.  Now although Graham (2007a)\footnote{See the end of his Appendix~A,
noting that the term $L^{0.5}$ should have read $L^{2}$.}, Graham \& Driver
(2007a, their section 3.2) and Graham (2008b, his section~2.2.2) warned that
the $M_{\rm bh}$--$L$ relation must be bent\footnote{At this time Bernardi et
al.\ (2007) had speculated that there may instead be curvature in the $M_{\rm
bh}$--$\sigma$ relation}, nobody had yet constructed this bent relation and it
was common practice to assume that the original relation, dominated by
luminous spheroids, could be extrapolated to low luminosities.  However in so doing, even
the classical, low-luminosity spheroids would end up incorrectly labeled
because they are offset from the extrapolated relation fit to the distribution of 
luminous spheroids in the $M_{\rm bh}$--$L$ diagram (Greene, Ho, \& Barth 2008; 
Jiang, Greene, \& Ho 2011; Kormendy, Bender, \& Cornell 2011). 
%

In a similar vein, and while on the discussion of pseudobulge identification,
a number of recent papers on the topic of supermassive black holes have
claimed the detection of pseudobulges because (due to their low S\'ersic index
$< 2$) they do not follow the extrapolations of other linear approximations to
the bright arms of what we now know are continuous but curved underlying
distributions. 
This important subject of non-(log-linear) scaling relations is extensively
reviewed in Graham (2012c, and references therein) and is vital if we are to
properly understand the formation of spheroids.  
For example, 
semi-analytical and theoretical models which still assume a log-linear
size-luminosity relation (e.g.\ Guo et al.\ 2010; Suh 2012) are handicapping
themselves as they are not matching galaxies in the real Universe.  
Galaxies which do not follow the bright arm of the curved (effective surface
brightness)--(effective radius) diagram should also not be labeled 
as pseudobulges for use in the various black hole diagrams 
(e.g.\ Gadotti \& Kauffmann 2009; Greene et al.\ 2008). 

Finally, we offer the following thoughts on whether or not barred galaxies 
might be offset in the $M_{\rm bh}$--$L$ diagram. 
Among lenticular disk galaxies, the average luminosity of the barred S0s is
apparently $\sim$0.4 mag fainter than that of the non-barred S0s, largely
because there are almost no barred S0 galaxies brighter than $M_B = -21$ mag
(van den Bergh 2012).  It has additionally been reported that the average
photometric bulge mass of barred galaxies is less than that of non-barred disc
galaxies of similar total galaxy stellar mass (Aguerri et al.\ 2005;
Laurikainen et al.\ 2007; Coelho \& Gadotti 2011, but see Laurikainen et al.\
2010), and that the bulge-to-total flux ratio is smaller in barred galaxies
than in non-barred galaxies of the corresponding Hubble type (Weinzirl et al.\
2009). If the effective sizes ($R_{\rm e}$) of bulges in barred galaxies are
also smaller than those in non-barred disk galaxies, then one may find that
the offset of barred galaxies toward higher velocity dispersions in the
$M_{\rm bh}$--$\sigma$ diagram is counter-balanced when using $\sigma^2R_{\rm
e}$ for bulge mass in the $M_{\rm bh}$--$M_{\rm spheroid}$ diagram.  Thus, rather
than finding barred galaxies tending to reside, on average, `rightward' of the
$M_{\rm bh}$--$L$ relation defined by the non-barred S\'ersic galaxies, the
bulges of barred galaxies may yet be found to reside `leftward' of it, i.e.\
'above' it.
Trying to discern if spheroids, and their black holes, in barred galaxies are
offset from those of non-barred galaxies in the $M_{\rm bh}$--$L$ diagram,
which was the 
question posed by Graham (2008a), will require either a larger data sample
than we currently have or reliable individual bulge/disc/bar decompositions.



\section{Conclusions}

Using a refined sample of 72 galaxies, 
comprised of 24 core-S\'ersic galaxies plus 48 S\'ersic galaxies, 
we have revised the $M_{\rm bh}$--$L_{\rm spheroid}$ relation. 
We have replaced what was a single power-law relation for all spheroids with 
distinct relations for the core-S\'ersic and the S\'ersic population.  

Using this exact same galaxy sample we have also updated the $M_{\rm
bh}$--$\sigma$ relations.  From 21 barred and 51 non-barred galaxies, we
confirm a mean factor of two offset between their $M_{\rm bh}$--$\sigma$
relations, and a steep slope of around 5 to 5.5 for the elliptical and
un-barred galaxies (see Figure~\ref{Fig2}a and Table~\ref{Tab3}).

It remains unknown if barred galaxies are offset, 
to lower black hole masses or equivalently brighter luminosities, 
relative to the non-barred S\'ersic galaxies, 
in the $M_{\rm bh}$--$L_{\rm spheroid}$ diagram. 

The lower mass S\'ersic galaxies are observed to be offset from $M_{\rm
bh}$--$L_{\rm spheroid}$ relation defined by the luminous galaxies.  However
rather than being an offset population of alleged pseudobulges, we have
explained why it is instead the luminous core-S\'ersic galaxies that deviate
from the $M_{\rm bh}$--$L_{\rm spheroid}$ relation defined by the fainter
S\'ersic galaxies. 

Luminous core-S\'ersic galaxies roughly follow the relation $M_{\rm bh}
\propto L^{1}$, consistent with the concept of their formation through dry
galaxy merging.  Built by such simple additive merger 
events, rather than what were once gaseous mechanisms, a roughly constant 
(black hole)-to-(host spheroid stellar) mass ratio of 0.35--0.55\% is observed.  
The fainter (barred and non-barred) S\'ersic galaxies follow an $M_{\rm
bh}$--$L_{\rm spheroid}$ relation which is some 2.5 times steeper.  Thus,
while the black hole mass to spheroid stellar mass ratio can be 0.5\% in
massive elliptical galaxies, it drops to values of 0.05\% in spheroids that 
are not much fainter than $M_B\approx -19$ mag.

We have discussed the implications of this steep S\'ersic $M_{\rm
bh}$--$L_{\rm spheroid}$ relation for the S\'ersic $M_{\rm bh}$--$M_{\rm
Sph,dyn}$ relation (Graham 2012a), which suggest that black holes grew
quadratically relative to their host spheroids.  We have therefore advocated a
new {\it quadratic} cold-gas `quasar' mode feeding equation (Eq.~\ref{EqQ}) 
for use in semi-analytical models.

Finally, using our new $M_{\rm bh}$--$L_{\rm spheroid}$ relation for S\'ersic
galaxies, we have predicted a number of intermediate mass black holes,
potentially filling a substantial portion of the current observational gap
between stellar mass black holes and one million solar mass black holes.

\section{acknowledgment}

This research was supported under the Australian Research Council’s
funding scheme (DP110103509 and FT110100263).

We acknowledge use of the HyperLeda database (http://leda.univ-lyon1.fr).
This research has made use of the NASA/IPAC Extragalactic Database (NED).


\begin{references}
 \reference{Aguer}Aguerri J.A.L., Iglesias-P\'aramo J., V\'{\i}lchez J.M., Munoz-Tun\'on C., \& S\'anchez-Janssen R.\ 2005b, AJ, 130, 475
 \reference{AaB96}Akritas, M.G., Bershady, M.A.\ 1996, ApJ, 470, 706
 \reference{Atk05}Atkinson, J.W., Collett, J.~L., Marconi, A., et al.\ 2005, MNRAS, 359, 504
 \reference{Bacon}Bacon, R., Emsellem, E., Combes, F., Copin, Y., Monnet, G., \& Martin, P.\ 2001, A\&A, 371, 409
 \reference{Bal7a}Balcells, M., Graham, A., \& Peletier, R.F.\ 2007, ApJ, 665, 1084
 \reference{BGH05}Barth, A.J., Greene, J.E., \& Ho, L.C.\ 2005, ApJ, 619, L151 
 \reference{Bart1}Barth, A.J., Sarzi, M., Rix, H.-W., Ho, L.C., Filippenko, A.V., \& Sargent, W.L.W.\ 2001, ApJ, 555, 685
 \reference{Bart4}Barth, A.J., Ho, L.C., Rutledge, R.E., Sargent, W.L.W.\ 2004, ApJ, 607, 90
 \reference{Bat10}Batcheldor D., 2010, ApJ, 711, L108
 \reference{Bei11}Beifiori, A., Courteau, S., Corsini, E.M., \& Zhu, Y.\ 2012, MNRAS, 419, 2497
 \reference{BBR80}Begelman, M.C., Blandford, R.D., \& Rees, M.J.\ 1980, Nature, 287, 307
 \reference{Bekki}Bekki, K., Couch, W.~J., Drinkwater, M.~J., \& Gregg, M.~D.\ 2001, ApJ, 557, L39
 \reference{DdJ01}Bell, E. F., \& de Jong, R. S.\ 2001, ApJ, 550, 212
 \reference{Ben88}Bender R., Doebereiner S., Moellenhoff C.\ 1988, A\&AS, 74, 385
 \reference{Ben05}Bender, R., Kormendy, J., Bower, G., et al.\ 2005, ApJ, 631, 280
 \reference{Ber07}Bernardi, M., Hyde, J.B., Sheth, R.K., Miller, C.J., \& Nichol, R.C.\ 2007, AJ, 133, 1741
 \reference{BST84}Binggeli, B., Sandage, A., \& Tarenghi, M.\ 1984, AJ, 89, 64
 \reference{Bo12a}Bogd{\'a}n, {\'A}., Forman, W.R., Zhuravleva, I., et al.\ 2012a, ApJ, 753, 140
 \reference{Bo12b}Bogd{\'a}n, {\'A}., Forman, W.R., Kraft, R.P., et al.\ 2012b, ApJ, 755, 25
 \reference{BMC13}Bonoli, S., Mayer, L., Callegari, S.\ 2012, MNRAS, submitted (arXiv:1211.3752)
 \reference{Bot05}Botte, V., Ciroi, S., di Mille, F., Rafanelli, P., Romano, A.\ 2005, MNRAS, 356, 789
 \reference{Bow01}Bower, G.A., Green, R.~F., Bender, R., et al.\ 2001, ApJ, 550, 75 
 \reference{Bla09}Blakeslee, J.P., Jord\'an, A., Mei, S., et al.\ 2009, ApJ, 694, 556
 \reference{Blake}Blakeslee, J.P., Lucey, J.R., Tonry, J.L., Hudson, M.J., Narayanan, V.K., \& Barris, B.J., 2002, MNRAS, 330, 443
 \reference{Blom2}Blom, C., Forbes, D.A., Brodie, J.P., et al.\ 2012, MNRAS, 426, 1959
 \reference{Bloom}Bloom, J.S., Giannios, D., Metzger, B.D., et al.\ 2011, Science, 333, 203
 \reference{Capet}Capetti, A., Marconi, A., Macchetto, D., \& Axon, D.\ 2005, A\&A, 431, 465
 \reference{Cap02}Cappellari, M., Verolme, E.K., van der Marel, R.P., Kleijn, G.A.V., Illingworth, G.D., Franx, M., Carollo, C.M., \& de Zeeuw, P.T.\ 2002, ApJ, 578, 787
 \reference{Cap06}Cappellari, M., Bacon, R., Bureau, M., et al.\ 2006, MNRAS, 366, 1126 
 \reference{Cap08}Cappellari, M., Bacon, R., Davies, R.~L., et al.\ 2008, in the proceedings of IAU Symposium 245, ``Formation and Evolution of Galaxy Bulges'', eds.\ M.Bureau, E.Athanassoula, and B.Barbuy, p.215
 \reference{Cap11}Cappellari, M., Emsellem, E., Krajnovi\'c, D., et al.\ 2011, MNRAS, 413, 813
 \reference{Car78}Carter, D.\ 1978, MNRAS, 182, 797
 \reference{Car87}Carter, D.\ 1987, ApJ, 312, 514
 \reference{Cen12}Cen, R.\ 2012, ApJ, 755, 28
 \reference{Cec01}Cecil, G., Bland-Hawthorn, J., Veilleux, S., Filippenko, A.V.\ 2001, ApJ, 555, 338
 \reference{Chi10}Chilingarian, I.V., Melchior, A.-L., \& Zolotukhin, I.Y.\ 2010, MNRAS, 405, 1409
 \reference{Cid04}Cid Fernandes, R., Gu, Q., Melnick, J., et al.\ 2004, MNRAS, 355, 273
 \reference{Ciras}Cirasuolo, M., Shankar, F., Granato, G.L., De Zotti, G., Danese, L.\ 2005, ApJ, 629, 816
 \reference{Cis11}Cisternas, M., Jahnke, K., Bongiorno, A., et al.\ 2011, ApJ, 741, L11
 \reference{CaG11}Coelho, P., \& Gadotti, D.A.\ 2011, ApJ, 743, L13
 \reference{Copin}Copin, Y., Cretton, N., \& Emsellem, E.\ 2004, A\&A, 415, 889
 \reference{Corbi}Corbin, M.R., O'Neil, E., \& Rieke, M.J.\ 2001, AJ, 121, 2549
 \reference{Cox00}Cox, D.P.\ 2000, Allen's Astrophysical quantities, New York: AIP Press; Springer
 \reference{CaVdB}Cretton, N., \& van den Bosch, F.\ 1999, ApJ, 514, 704
 \reference{Croto}Croton, D.~J., Springel, V., White, S.~D.~M., et al.\ 2006, MNRAS, 365, 11
 \reference{Dalla}Dalla Bont\`a E., Ferrarese L., Corsini E.M., Miralda-Escud\'e J., Coccato L., Sarzi M., Pizzella A., \& Beifiori A.\ 2009, ApJ, 690, 537
 \reference{DEF83}Davies, R.L., Efstathiou, G., Fall, S.M., et al.\ 1983, ApJ, 266, 41
 \reference{Dav06}Davies, R.I., Thomas, J., Genzel, R., et al.\ 2006, ApJ, 646, 754 
 \reference{DMW12}DeBuhr, J., Ma, C.-P., \& White, S.D.M.\ 2012, MNRAS, 426, 983
 \reference{Dec08}Decarli, R., Dotti, M., Fontana, M., Haardt, F., 2008, MNRAS, 386, L15
 \reference{deFra}de Francesco, G., Capetti, A., \& Marconi, A.\ 2006, A\&A, 460, 439
 \reference{FCM08}de Francesco, G., Capetti, A., Marconi, A.\ 2008, A\&A, 479, 355
 \reference{deG12}DeGraf, C., Di Matteo, T., Khandai, N., \& Croft, R.\ 2012, ApJ, 755, L8
 \reference{deR05}de Rijcke, S., Michielsen, D., Dejonghe, H., Zeilinger, W.W., \& Hau, G.K.T.\ 2005, A\&A, 438, 491
 \reference{Des12}Deshmukh, S.P., Tate, B.T., Vagshette, N.D., Pandey, S.K., \& Patil, M.K.\ 2012, submitted to RAA (arXiv:1207.4324)
 \reference{Des07}Desroches, L.-B., Quataert, E., Ma, C.-P., \& West, A.~A.\ 2007, MNRAS, 377, 402
 \reference{deV91}De Vaucouleurs, G., De Vaucouleurs, A., Corwin, H.~G.~Jr, Buta R.~J., Paturel G., \& Fouque P.\ 1991, Third Reference Catalogue of Bright Galaxies. Springer-Verlag, Berlin (RC3)
 \reference{deVOl}de Vaucouleurs, G. \& Olson, D. W.\ 1982, ApJ, 256, 346
 \reference{Dev03}Devereux, N.A., Ford, H.C., Tsvetanov, Z., \& Jacoby, G.\ 2003, AJ, 125, 1226
 \reference{DaD87}Djorgovski, S., \& Davis, M.\ 1987, ApJ, 313, 59
 \reference{Dong2}Dong, X.-B., Ho, L.C., Yuan, W., et al.\ 2012, ApJ, 755, 167
 \reference{DaD06}Dong, X.Y., \& De Robertis, M.M.\ 2006, AJ, 131, 1236
 \reference{Dri08}Driver, S.~P., Popescu, C.~C., Tuffs, R.~J., Graham, A.W., Liske, J., \& Baldry, I.\ 2008, ApJ, 678, L101
 \reference{Dub12}Dubois, Y., Devriendt, J., Slyz, A., \& Teyssier, R.\ 2012, MNRAS, 420, 2662
 \reference{DaG12}Dullo, B., \& Graham, A.W.\ 2012, ApJ, submitted
 \reference{EMO91}Ebisuzaki, T., Makino, J., \& Okumura, S.K.\ 1991, Nature, 354, 212
 \reference{Ems11}Emsellem, E., Cappellari, M., Krajnovi\'c, D., et al.\ 2011, MNRAS, 414, 888
 \reference{EDB99}Emsellem, E., Dejonghe, H., \& Bacon, R.\ 1999, MNRAS, 303, 495
 \reference{Fab87}Faber, S.M., Dressler, A., Davies, R.L., Burstein, D.,
 Lynden-Bell, D., Terlevich, R., \& Wegner, G.\ 1987, in Nearly Normal
 Galaxies: From the Planck Time to the Present, edited by S.M.Faber (Springer,
 New York), p.175
 \reference{Erw3}Erwin, P., Beltr\'an, J.C.V., Graham, A.W., \& Beckman, J.E.\ 2003, ApJ, 597, 929
 \reference{FaJ76}Faber, S.M. \& Jackson, R.E.\ 1976, ApJ, 204, 668 
 \reference{Fab97}Faber, S.M., Tremaine, S., Ajhar, E.A., et al.\ 1997, AJ, 114, 1771
 \reference{FKM04}Falcke, H., K\"ording, E., Markoff, S.\ 2004. A\&A, 414, 895
 \reference{F-B11}Falc\'on-Barroso, J., van de Ven, G., Peletier, R.F., et al.\ 2011, MNRAS, 417, 1787
 \reference{Far09}Farrell, S.A., Webb, N.A., Barret, D., Godet, O., \& Rodrigues, J.M.\ 2009, Nature, 460, 73
 \reference{Fer96}Ferrarese, L., Freedman, W.~L., Hill, R.~J., et al.\ 1996, ApJ, 464, 568
 \reference{Fer02}Ferrarese, L.\ 2002, ApJ, 578, 90
 \reference{Ferra}Ferrarese, L., C\^ot\'e, P., Jord\'an, A., et al.\ 2006, ApJS, 164, 334 
 \reference{FaF99}Ferrarese, L., \& Ford, H.C.\ 1999, ApJ, 515, 583
 \reference{FaF05}Ferrarese, L., \& Ford, H.C.\ 2005, Space Science Reviews, 116, 523 
 \reference{FaM00}Ferrarese, L.A., \& Merritt, D.\ 2000, ApJ, 539, L9
 \reference{Fer94}Ferrarese, L., van den Bosch, F.C., Ford, H.C., Jaffe, W., \& O'Connell, R.W.\ 1994, AJ, 108, 1598 
 \reference{FaH03}Filippenko, A.V., \& Ho, L.C.\ 2003, ApJ, 588, L13
 \reference{FaD08}Fisher, D.B., Drory, N.\ 2008, AJ, 136, 773
 \reference{FKJ12}For, B.-Q., Koribalski, B., Jarrett, T.\ 2012, MNRAS, 425, 1934
 \reference{Freed}Freedman, W.~L., Madore, B.~F., Gibson, B.~K., et al.\ 2001, ApJ, 553, 47
 \reference{GdK09}Gadotti D.A., \& Kauffmann G.\ 2009, MNRAS, 399, 621
 \reference{Gav05}Gavazzi, G., Donati, A., Cucciato, O., Sabatini, S., Boselli, A., Davies, J., \& Zibetti, S.\ 2005, A\&A, 430, 411
 \reference{Geb11}Gebhardt, K., Adams, J., Richstone, D., et al.\ 2011, ApJ, 729, 119 
 \reference{Geb03}Gebhardt, K., Richstone, D., Tremaine, S., et al.\ 2003, ApJ, 583, 92
 \reference{Geb00}Gebhardt, K., Bender, R., Bower, G., et al.\ 2000, ApJ, 539, L13 
 \reference{Geb07}Gebhardt, K., Lauer, T.~R., Pinkney, J., et al.\ 2007, ApJ, 671, 1321
 \reference{GKM02}Gerssen, J., Kuijken, K., \& Merrifield, M.R.\ 1999, MNRAS, 306, 926
 \reference{Gez12}Gezari, S., Chornock, R., Rest, A., et al.\ 2012, Nature, 485, 217
 \reference{Gil09}Gillessen, S., Eisenhauer, F., Fritz, T.K., Bartko, H., Dodds-Eden, K., Pfuhl, O., Ott, T., Genzel, R.\ 2009, ApJ, 707, L114
 \reference{Gonz8}Gonz\'alez Delgado, R.M., P\'erez, E., Cid Fernandes, R., \& Schmitt, H.\ 2008, AJ, 135, 747
 \reference{Gra04}Graham, A.W.\ 2002, ApJ, 568, L13
 \reference{Gra04}Graham, A.W.\ 2004, ApJ, 613, L33
 \reference{Gra7a}Graham, A.W.\ 2007b, BAAS, 39 759 \#13.27
 \reference{Gra7b}Graham, A.W.\ 2007a, MNRAS, 379, 711
 \reference{Gra8a}Graham, A.W.\ 2008a, ApJ, 680, 143
 \reference{Gra8b}Graham, A.W.\ PASA, 2008b, 25, 167
 \reference{Gr11a}Graham, A.W.\ 2011a, in ``A Universe of dwarf galaxies'', EAS Pub.\ Ser., 48, 231 (arXiv:1009.5002)
 \reference{Gr11b}Graham, A.W.\ 2011b, (arXiv:1103.0525)
 \reference{Gr12a}Graham, A.W.\ 2012a, ApJ, 746, 113
 \reference{Gr12b}Graham, A.W.\ 2012b, MNRAS, 422, 1586
 \reference{Gr12c}Graham, A.W.\ 2012c, in ``Planets, Stars and Stellar Systems'', Springer Publishing (arXiv:1108.0997)
 \reference{GaD05}Graham, A.W., \& Driver, S.P.\ 2005, PASA, 22(2), 118
 \reference{GaD7a}Graham, A.W., \& Driver, S.P.\ 2007a, ApJ, 655, 77
 \reference{GaD7b}Graham, A.W., \& Driver, S.P.\ 2007b, MNRAS, 380, L15
 \reference{Gra03}Graham, A.W., Erwin, P., Trujillo, I., \& Asensio Ramos, A.\ 2003, AJ, 125, 2951
 \reference{GaG03}Graham, A.W. \& Guzm\'an, R.\ 2003, AJ, 125, 2936
 \reference{GaL09}Graham, A.W., \& Li, I-H.\ 2009, ApJ, 698, 812
 \reference{Get11}Graham, A.W., Onken, C., Athanassoula, L., \& Combes, F.\ 2011, MNRAS. 412, 2211
 \reference{GaS09}Graham, A.~W., \& Spitler, L.~R.\ 2009, MNRAS, 397, 2148 
 \reference{GaW08}Graham, A.~W., \& Worley, C.~C.\ 2008, MNRAS, 388, 1708
 \reference{GHB08}Greene, J.E., Ho, L.C., \& Barth, A.J.\ 2008, ApJ, 688, 159
 \reference{Gre09}Greene, J.E., Zakamska, N.L., Liu, X., Barth, A.J., \& Ho, L.C.\ 2009, ApJ, 702, 441
 \reference{Gre10}Greene, J.E., Peng, C.Y., Kim, M., et al.\ 2010, ApJ, 721, 26 
 \reference{Gr97a}Greenhill, L.J., Moran, J.M., \& Herrnstein, J.R.\ 1997, ApJ, 481, L23
 \reference{Gre03}Greenhill, L.J., Booth, R.~S., Ellingsen, S.~P., et al.\ 2003, ApJ, 590, 162
 \reference{Grill}Grillmair, C.J., Faber, S.M., Lauer, T.R., et al.\ 1994, AJ, 108, 102
 \reference{Groot}Grootes, M.W., Tuffs, R.J., Popescu, C.C., et al.\ MNRAS, in press 
 \reference{Guk9a}G\"ultekin K., Richstone, D.~O., Gebhardt, K., et al.\ 2009a, ApJ, 695, 1577
 \reference{Guk9b}G\"ultekin K., Richstone, D.~O., Gebhardt, K., et al.\ 2009b, ApJ, 698, 198
 \reference{Gut11}G{\"u}ltekin, K., Richstone, D.O., Gebhardt, K., et al.\ 2011, ApJ, 741, 38
 \reference{Guo11}Guo, Q., White, S., Boylan-Kolchin, M., et al.\ 2011, MNRAS, 413, 101
 \reference{HaR04}H\"aring, N. \& Rix, H.-W.\ 2004, ApJ, 604, L89
 \reference{Hel92}Held, E. V., de Zeeuw, T., Mould, J., \& Picard, A.\ 1992, AJ, 103, 851
 \reference{Herr9}Herrnstein, J.R., Moran, J.~M., Greenhill, L.~J., et al.\ 1999, Nature, 400, 539
 \reference{Hernz}Hern{\'a}ndez-Toledo, H.~M., Cano-D{\'{\i}}az, M., Valenzuela, O., et al.\ 2011, AJ, 142, 182
 \reference{Her08}Herrmann, K. A., Ciardullo, R., Feldmeier, J. J., \& Vinciguerra, M. 2008, ApJ, 638, 630
 \reference{HaM08}Hicks, E.K.S., \& Malkan, M.A.\ 2008, ApJS, 174, 31
 \reference{Hiner}Hiner, K.D., Canalizo, G., Wold, M., Brotherton, M.S., \& Cales, S.L.\ 2012, ApJ, 756, 162
 \reference{Hirsc}Hirschmann, M., Somerville, R.S., Naab, T., Burkert, A.\ 2012, MNRAS, 426, 237
 \reference{Hla12}Hlavacek-Larrondo, J., Fabian, A.C., Edge, A.C., Hogan, M.T.\ 2012, MNRAS, 424, 224
 \reference{H0999}Ho, L.C.\ 1999, in Observational Evidence for Black Holes in the Universe, ed. S.K. Chakrabarti (Dordrecht: Kluwer), 157
 \reference{Ho007}Ho, L.C.\ 2007, ApJ, 668, 94
 \reference{HoL09}Ho, L.C., Greene, J.E., Filippenko, A.V., \& Sargent, W.L.W.\ 2009, ApJS, 183, 1
 \reference{Hop06}Hopkins, P.F., Hernquist, L., Cox, T.J., et al.\ 2006, ApJS, 163, 1
 \reference{H1399}Houghton, R.C.W., Magorrian, J., Sarzi, M., Thatte, N., Davies, R.L., \& Krajnovi\'c, D.\ 2006, MNRAS, 367, 2
 \reference{Hu008}Hu, J.\ 2008, MNRAS, 386, 2242
 \reference{Ish01}Ishihara, Y., Nakai, N., Iyomoto, N., Makishima, K., Diamond, P., \& Hall, P.\ 2001, PASJ, 53, 215
 \reference{Jaffe}Jaffe, W., Ford, H.C., O'Connell, R.W., van den Bosch, F.C., Ferrarese, L.\ 1994, AJ, 108, 1567
 \reference{JaM11}Jahnke, K., \& Macci{\`o}, A.~V.\ 2011, ApJ, 734, 92 
 \reference{Jar11}Jardel, J.R., Gebhardt, K., Shen, J., et al.\ 2011, ApJ, 739, 21 
 \reference{2MASS}Jarrett, T.H., Chester, T., Cutri, R., et al.\ 2000, AJ, 119, 2498 (2MASS)
 \reference{JBB04}Jerjen, H., Binggeli, B., \& Barazza, F.D.\ 2004, AJ, 127, 771
 \reference{JGH12}Jiang, Y.-F., Greene, J.E., \& Ho, L.C.\ 2011, ApJ, 737, L45
 \reference{Jim11}Jim\'enez, N., Cora, S.A., Bassino L.P., Tecce T.E., \& Smith Castelli A.V.\ 2011, MNRAS, 417, 785
 \reference{JNB09}Johansson, P.H., Naab, T., Burkert, A.\ 2009, ApJ, 690, 802
 \reference{Karac}Karachentsev, I.D., Tully, R.~B., Dolphin, A., et al.\ 2007, AJ, 133, 504
 \reference{KaH00}Kauffmann, G., \& Haehnelt, M.\ 2000, MNRAS, 311, 576
 \reference{Kha12}Khandai, N., Feng, Y., DeGraf, C., Di Matteo, T.,Croft, R.A.C.\ 2012, MNRAS, 423, 2397
 \reference{Kho12}Khorunzhev, G.A., Sazonov, S.Y., Burenin, R.A., Tkachenko, A.Y.\ 2012, Astronomy Letters, 38(8), 475
 \reference{Kin78}King I.R.\ 1978, ApJ, 222, 1
 \reference{KaK10}Kisaka, S., \& Kojima, Y.\ 2010, MNRAS, 405, 1285
 \reference{Knap3}Knapen, J.H., de Jong, R.S., Stedman, S., \& Bramich, D.M.\ 2003, MNRAS, 344, 527
 \reference{KaX07}Komossa S., Xu D., 2007, ApJ 667, L33
 \reference{KGM05}Kondratko, P.T., Greenhill, L.J., \& Moran, J.M.\ 2005, ApJ, 618, 618
 \reference{Kond8}Kondratko, P.T., Greenhill, L.J., \& Moran, J.M.\ 2008, ApJ, 678, 87
 \reference{Kor85}Kormendy, J.\ 1985, ApJ, 295, 73
 \reference{KBC11}Kormendy, J., Bender, R., \& Cornell, M.E.\ 2011, Nature, 469, 374
 \reference{Kor09}Kormendy, J., Fisher, D.B., Cornell, M.E., \& Bender, R.\ 2009, ApJS, 182, 216
 \reference{KaG01}Kormendy, J., \& Gebhardt, K., in H.Martel, J.C.Wheeler, eds, 2001 {\it 20th Texas Symposium on relativistic astrophysics}, Am. Inst. Phys., New York, Conf.\ Proc., 586, 363
 \reference{KaR95}Kormendy, J., \& Richstone, D.\ 1995, ARA\&A, 33, 581
 \reference{KMc09}Kornei, K.A., \& McCrady, N.\ 2009, ApJ, 697, 1180
 \reference{Kou12}Kourkchi, E., Khosroshahi, H. G., Carter, D., et al., 2012, MNRAS, 420, 2819
 \reference{Kraj9}Krajnovi{\'c} D., McDermid R.M., Cappellari M., \& Davies R.L.\ 2009, MNRAS, 399, 1839
 \reference{KSB87}Kunth, D., Sargent, W.L.W., Bothun, G.D.\ 1987, AJ, 93, 29
 \reference{Kuo11}Kuo, C.Y., Braatz, J.A., Condon, J.J., et al.\ 2011, ApJ, 727, 20 
 \reference{LaB10}La Barbera, F., de Carvalho, R.R., de La Rosa, I.G., \& Lopes, P.A.A.\ 2010, MNRAS, 408, 1335 
 \reference{Lahav}Lahav, C.G., Meiron, Y., Soker, N.\ 2011 (arXiv:1112.0782)
 \reference{Lai03}Laine, S., van der Marel, R.P., Lauer, T.R., et al.\ 2003, AJ, 125, 478 
 \reference{Laor1}Laor, A.\ 2001, ApJ, 553, 677
 \reference{Lau05}Lauer, T.R., Faber, S.~M., Gebhardt, K., et al.\ 2005, AJ, 129, 2138
 \reference{Lau07}Lauer, T.R., Faber, S.~M., Richstone, D., et al.\ 2007, ApJ, 662, 808
 \reference{Laur5}Laurikainen, E., Salo, H., \& Buta, R.\ 2005, MNRAS, 362, 1319
 \reference{Laur7}Laurikainen, E., Salo, H., Buta, R., \& Knapen, J.~H.\ 2007, MNRAS, 381, 401 
 \reference{Lau10}Laurikainen, E., Salo, H., Buta, R., Knapen, J.~H., \& Comer{\'o}n, S.\ 2010, MNRAS, 405, 1089
 \reference{Lauri}Laurikainen, E., Salo, H., Buta, R., \& Knapen, J.~H.\ 2011, MNRAS, 418, 1452
 \reference{LBK12}Leigh, N., B{\"o}ker, T., \& Knigge, C.\ 2012, MNRAS, 424, 2130
 \reference{Levan}Levan, A.J., Tanvir, N.R., Cenko, S.B., et al.\ 2011, Science, v.333, no.6039, p.199
 \reference{LHW11}Li, Y.-R., Ho, L.C., \& Wang, J.-M.\ 2011, ApJ, 742, 33
 \reference{LWH12}Li, Y.-R., Wang, J.-M., \& Ho, L.C.\ 2012, IAU Symp.\ 290, C.M.\ Zhang, T.\ Belloni, M.\ Mendez \& S.N.\ Zhang (eds.)
 \reference{Liu09}Liu, F.S., Xia, X.Y., Mao, S., Wu, H., \& Deng, Z.G.\ 2008, MNRAS, 385, 23
 \reference{LaB03}Lodato, G., \& Bertin, G.\ 2003, A\&A, 398, 517
 \reference{Lyn88}Lynden-Bell D., Faber S.M., Burstein D., Davies R.L., Dressler A., Terlevich R.J., \& Wegner G.\ 1988, ApJ, 326, 19
 \reference{Mach4}Macchetto, F., Capetti, A., Sparks, W.B., Axon, D.J., \& Boksenberg, A.\ 1994, ApJ, 435, L15
 \reference{Mad99}Madore B.F., Freedman, W.~L., Silbermann, N., et al., 1999, ApJ, 515, 29
 \reference{Mag98}Magorrian, J., Tremaine, S., Richstone, D., et al., 1998, AJ, 115, 2285
 \reference{Mag12}Magoulas, C., Springob, C.M., Colless, M.M., et al.\ 2012, MNRAS, in press (arXiv:1206.0385)
 \reference{MGT98}Malkan, M.A., Gorjian, V., \& Tam, R. 1998, ApJS, 117, 25
 \reference{MaK81}Malumuth, E.M., \& Kirshner. R.P.\ 1981, ApJ, 251, 508
 \reference{Map12}Mapelli M., Ripamonti E., Vecchio A., Graham A.W., \& Gualandris A.\ 2012, A\&A, 542, 102
 \reference{Marc8}Marconi, A., Axon, D.J., Maiolino, R., Nagao, T., Pastorini, G., Pietrini, P., Robinson, A., Torricelli, G.,\ 2008, ApJ, 678, 693
 \reference{MaH03}Marconi, A., \& Hunt, L.K.\ 2003, ApJ, 589, L21
 \reference{Mar00}Marconi, A., Oliva, E., van der Werf, P.P., et al.\ 2000, A\&A, 357, 24
 \reference{Mat12}Mathur, S., Fields, D., Peterson, B.M., \& Grupe, D.\ 2012, ApJ, 754, 146
 \reference{MaG05}Mathur S., Grupe D., 2005, ApJ, 633, 688
 \reference{MaG03}Matkovi\'c, A., \& Guzm\'an, R.\ 2005, MNRAS, 362, 289
 \reference{McA11}McAlpine, W., Satyapal, S., Gliozzi, M., et al.\ 2011, ApJ, 728, 25
 \reference{Mc11a}McConnell, N.J., Ma, C.-P., Gebhardt, K., et al.\ 2011a, Nature, 480, 215
 \reference{Mc11b}McConnell, N.J., Ma, C.-P., Graham, J.R., et al.\ 2011b, ApJ, 728, 100 
 \reference{McC12}McConnell, N.J., Ma, C.-P., Murphy J.D., et al.\ 2012, ApJ, 756 179 
 \reference{MaD02}McLure R. J., \& Dunlop J. S.\ 2002, MNRAS, 331, 795
 \reference{MaD04}McLure R. J., \& Dunlop J. S.\ 2004, MNRAS, 352, 1390 
 \reference{Mei07}Mei, S., Blakeslee, J.P., \& C\^ot\'e, P., et al.\ 2007, ApJ, 655, 144
 \reference{MHdi3}Merloni, A., Heinz, S., \& di Matteo, T.\ 2003, MNRAS, 345, 1057
 \reference{MaK99}Merrifield, M.R., \& Kuijken, K.\ 1999, A\&A, 345, L47
 \reference{MaF1a}Merritt D., Ferrarese L., 2001a, ApJ, 547, 140
 \reference{MaF1b}Merritt D., \& Ferrarese L.\ 2001b, in “The Central Kiloparsec of Starbursts and AGN: The La Palma Connection”, ASP Conference Proceedings Vol.\ 249, J.H.\ Knapen, J.E.\ Beckman, I.\ Shlosman, and T.J.\ Mahoney eds., San Francisco: Astronomical Society of the Pacific, p.335.
 \reference{MMS07}Merritt, D., Mikkola, S., \& Szell, A.\ 2007, ApJ, 671, 53 
 \reference{Mun07}Mu\~noz Mar{\'{\i}}n, V.M., Gonz\'alez Delgado, R.M., Schmitt, H.R., et al.\ 2007, AJ, 134, 648
 \reference{Nak98}Nakai, N., Inoue, M., Hagiwara, Y., Miyoshi, M., \& Diamond, P.J.\ 1998, IAU Colloq.~164: Radio Emission from Galactic and Extragalactic Compact Sources, 144, 237
 \reference{NaW95}Nelson, C.H., \& Whittle, M.\ 1995, ApJS, 99, 67
 \reference{Neu10}Neumayer, N., 2010, Pub.\ Astron.\ Soc.\ Aust., 27, 449
 \reference{NaW12}Neumayer, N., \& Walcher, C.J.\ 2012, Advances in Astronomy, 2012, id.\ 709038
 \reference{Nieto}Nieto, J.-L., Bender, R., \& Surma, P.\ 1991, A\&A, 244, L37
 \reference{Now08}Nowak, N., Saglia, R.P., Thomas, J., Bender, R., Davies R.I., \& Gebhardt K.\ 2008, MNRAS, 391, 1629
 \reference{Nowak}Nowak, N., Saglia, R.P., Thomas, J., Bender, R., Pannella, M., Gebhardt, K., \& Davies, R.I.\ 2007, MNRAS, 379, 909
 \reference{Now10}Nowak N., Thomas J., Erwin P., Saglia R.P., Bender R., \& Davies R.I.\ 2010, MNRAS, 403, 646
 \reference{Nyl12}Nyland, K., Marvil, J., Wrobel, J.M., Young, L.M., \& Zauderer, B.A.\ 2012, ApJ, 753, 103 
 \reference{OOKM5}Oliva E., Origlia L., Kotilainen J.K., \& Moorwood A.F.M.\ 1995, A\&A 301, 55
 \reference{Onk04}Onken, C.A., Ferrarese, L., Merritt, D., et al.\ 2004, ApJ, 615, 645
 \reference{Onken}Onken, C.A., Valluri, M., Peterson, B.~M., et al.\ 2007, ApJ, 670, 105
 \reference{Park2}Park, D., Kelly, B.C., Woo, J.-H., Treu, T.\ 2012, ApJS, 203, 6
 \reference{Pas07}Pastorini, G., Marconi, A., Capetti, A., et al.\ 2007, A\&A, 469, 405
 \reference{Pat03}Paturel G., Petit C., Prugniel P., Theureau G., Rousseau J., Brouty M., Dubois P., \& Cambr{\'e}sy L.\ 2003, A\&A, 412, 45
 \reference{Pel90}Peletier, R.F., Davies, R.L., Illingworth, G.D., Davis, L.E., Cawson, M.\ 1990, AJ, 100, 1091
 \reference{Pen07}Peng, C.-Y.\ 2007, ApJ, 671, 1098
 \reference{Pet05}Peterson, B.M., Bentz, M.C., Desroches, L.-B., et al.\ 2005, ApJ, 632, 799
 \reference{PaM02}Pogge, R.W., \& Martini, P.\ 2002, ApJ, 569, 624
 \reference{Pop00}Popescu, C.C., Misiriotis, A., Kylafis, N.D., Tuffs, R.J., \& Fischera, J.\ 2000, A\&A, 362, 138
 \reference{Por12}Portinari, L., Kotilainen, J., Falomo, R., \& Decarli, R.\ 2012, MNRAS, 420, 732
 \reference{Quill}Quillen, A.~C., Bower, G.~A., \& Stritzinger, M.\ 2000, ApJS, 128, 85 
 \reference{Rad08}Radomski, J.T., Packham, C., Levenson, N.A., et al.\ 2008, ApJ, 681, 141
 \reference{Rav01}Ravindranath, S., Ho, L., Peng, C., et al.\ 2001, AJ, 122, 653
 \reference{Rek05}Rekola R., Richer M.G., McCall M.L., Valtonen M.J., Kotilainen J.K., \& Flynn C.\ 2005, MNRAS, 361, 330
 \reference{Rest1}Rest, A., van den Bosch, F.C., Jaffe, W., et al.\ 2001, 121, 2431
 \reference{Rich1}Richings, A.J., Uttley, P. and K\'ording, E.\ 2011, MNRAS, 415, 2158
 \reference{Ric98}Richstone, D., Ajhar, E.A., Bender, R., et al.\ 1998, Nature, 395, A14
 \reference{Rix99}Rix, H.-W., Carollo, C.M., \& Freeman, K.\ 1999, ApJ, 513, L25
 \reference{Rod06}Rodr{\'{\i}}guez-Rico, C.A., Goss, W.M., Zhao, J.-H., G{\'o}mez, Y., \& Anantharamaiah, K.R.\ 2006, ApJ, 644, 914
 \reference{Rus11}Rusli, S.P., Thomas, J., Erwin, P., et al.\ 2011, MNRAS, 410, 1223 
 \reference{SaM96}Sanders, D.B., \& Mirabel, I.F.\ 1996, ARA\&A, 34, 749
 \reference{San11}Sani, E., Marconi, A., Hunt, L.K., \& Risaliti, G.\ 2011, MNRAS, 413, 1479
 \reference{Sarzi}Sarzi, M., Rix, H.-W., Shields, J.C., Rudnick, G., Ho, L.C., McIntosh, D.H., Filippenko, A.V., \& Sargent, W.L.W.\ 2001, ApJ, 550, 65
 \reference{Sch80}Schechter, P.L.\ 1980, AJ, 85, 801
 \reference{SaF11}Schlafly, E.F., \& Finkbeiner, D.P.\ 2011, ApJ, 737, 103 
 \reference{SFD98}Schlegel, D.J., Finkbeiner, D.P., \& Davis, M.\ 1998, ApJ, 500, 525 
 \reference{Sch11}Schombert, J., 2011, arXiv:1107.1728
 \reference{SaS12}Schombert, J., \& Smith, A.K.\ 2012, PASA, PASA, 29, 174 
 \reference{SaW11}Schulze, A., \& Wisotzki, L.\ 2011, A\&A, 535, A87
 \reference{SaG12}Scott, N., \& Graham, A.W.\ 2012, ApJ, submitted (arXiv:1205.5338)
 \reference{Ser63}S\'ersic, J.-L.\ 1963, Boletin de la Asociacion Argentina de Astronomia, vol.6, p.41
 \reference{Ses12}Sesana, A.\ 2012, Advances in Astronomy, vol.\ 2012, id.\#805402 (arXiv:1110.6445) 
 \reference{Set08}Seth, A., Ag\"ueros, M., Lee, D., Basu-Zych, A.\ 2008, 678, 116
 \reference{Set12}Seth, A.C., Cappellari, M., Neumayer, N., et al.\ 2010, ApJ, 714, 713
 \reference{Sey12}Seymour, N., Altieri, B., De Breuck, C., et al.\ 2012, ApJ, 755, 146
 \reference{Shank}Shankar, F., Marulli, F., Mathur, S., Bernardi, M., \& Bournaud, F.\ 2012, A\&A, 540, A23
 \reference{Sha93}Shaw, M.A., Wilkinson, A., \& Carter, D.\ 1993, A\&A, 268, 511
 \reference{SaJ09}Shen, J., \& Gebhardt, K.\ 2010, ApJ, 711, 484
 \reference{Shi08}Shields, J.C., Walcher, C.J., B{\"o}ker, T., et al.\ 2008, ApJ, 682, 104
 \reference{Smith}Smith, R.J., Lucey, J.R., Hudson, M.J., Schlegel, D.J., \& Davies R.L.\ 2000, MNRAS, 313, 469
 \reference{Soria}Soria, R., Hau, G.K.T., Graham, A.W., Kong, A.K.H., Kuin, N.P.M., Li, I.-H., Liu, J.-F. \& Wu, K.\ 2010, MNRAS, 405, 870
 \reference{Suh12}Suh, P.K.\ 2012, International Journal of Astronomy and Astrophysics, 2, 101
 \reference{Sut12}Sutton, A. D., Roberts, T. P., Walton, D. J., Gladstone, J.C., Scott, A.E., 2012, MNRAS, 423, 1154
 \reference{Tad03}Tadhunter, C., Marconi, A., Axon, D., Wills, K., Robinson, T.G., \& Jackson, N.\ 2003, MNRAS, 342, 861
 \reference{Ton81}Tonry, J.\ 1981, ApJ, 251, L1
 \reference{Ton01}Tonry, J.~L., Dressler, A., Blakeslee, J.~P., et al.\ 2001, ApJ, 546, 681
 \reference{Tre04}Tremonti, C.A., Heckman, T.M., Kauffmann, G., et al.\ 2004, ApJ, 613, 898 
 \reference{Trott}Trotter, A. S., Greenhill, L. J., Moran, J. M., Reid, M. J., Irwin, J. A., \& Lo, K. 1998, ApJ, 495, 740
 \reference{Tru02}Trujillo, I., Asensio Ramos, A., Rubi\~no-Martin, J.A., Graham, A.W., Aguerri, J.A.L., Cepa, J., Guti\'errez, C.M.\ 2002, MNRAS, 333, 510
 \reference{Tru04}Trujillo, I., Erwin, P., Asensio Ramos, A., \& Graham, A.W.\ 2004, AJ, 127, 1917
 \reference{Tuf04}Tuffs, R.J., Popescu, C.C., V\"olk, H.J., Kylafis, N.D., \& Dopita, M.A.\ 2004, A\&A, 419, 821
 \reference{VME04}Valluri, M., Merritt, D., \& Emsellem, E.\ 2004, ApJ, 602, 66
 \reference{VCL12}Valtonen, M.J., Ciprini, S. \& Lehto, H.J.\ 2012, MNRAS, 427, 77
 \reference{vdB12}van den Bergh, S.\ 2012, ApJ, 754, 68
 \reference{vaZ09}van den Bosch R.C.E., \& de Zeeuw P.T.\ 2010, MNRAS, 401, 1770 
 \reference{VjV98}van den Bosch, F.C., Jaffe, W., \& van der Marel, R.P.\ 1998, MNRAS, 293, 343
 \reference{vav98}van der Marel, R.P., \& van den Bosch, F.C.\ 1998, AJ, 116, 2220
 \reference{Vau05}Vaughan, S., Iwasawa, K., Fabian, A.C., Hayashida, K.\ 2005, MNRAS, 356, 524
 \reference{VBC99}Veilleux, S., Bland-Hawthorn, J., \& Cecil, G.\ 1999, AJ, 118, 2108
 \reference{Veret}Verolme, E.K., et al.\ 2002, MNRAS, 335, 517
 \reference{Vik11}Vika, M., Driver, S. P., Cameron, E., Kelvin, L., \& Robotham, A.\ 2011, MNRAS, 419, 2264
 \reference{Vik09}Vika, M., Driver, S., Graham, A., \& Liske, J.\ 2009, MNRAS, 400, 1451
 \reference{MVM12}Villar-Mart\'{\i}n, M., Cabrera Lavers, A., Bessiere, P., Tadhunter, C., Rose, M. \& de Breuck, C.\ 2012, MNRAS, 423, 80
 \reference{Vio12}Violette Impellizzeri, C.M., Braatz, J.A., Kuo, C.-Y., Reid,
 M.J., Lo, K.Y., Henkel, C., \& Condon, J.J.\ 2012, in ``Cosmic Masers- from
 OH to H0'', Cambridge University Press, R.S. Booth, E.M.L. Humphreys \&
 W.H.T. Vlemmings, eds., IAU Symposium 287, 311
 \reference{VDL07}von der Linden, A., Best, P.N., Kauffmann, G., \& White, S.D.M.\ 2007, 379, 867
 \reference{Walsh}Walsh, J.L., Barth, A.J., \& Sarzi, M.\ 2010, ApJ, 721, 762 
 \reference{Walsh}Walsh, J.L., van den Bosch, R.C.E., Barth, A.J., et al.\ 2012, ApJ, 753, 79 
 \reference{Wan02}Wandel A., 2002, ApJ, 565, 762
 \reference{Weinz}Weinzirl, T., Jogee, S., Khochfar, S., et al.\ 2009, ApJ, 696, 411 
 \reference{Wold6}Wold, M., Lacy, M., K{\"a}ufl, H.U., \& Siebenmorgen, R.\ 2006, A\&A, 460, 449
 \reference{WaG06}Wold, M., \& Galliano, E.\ 2006, MNRAS, 369, L47
 \reference{Wu009}Wu Q., 2009, MNRAS, 398, 1905
 \reference{Yama4}Yamauchi, A., Nakai, N., Sato, N., \& Diamond, P.\ 2004, PASJ, 56, 605
 \reference{YaW75}Yoshizawa, M., \& Wakamatsu, K.\ 1975, A\&A, 44, 363
  \reference{ZLY12}Zhang, X., Lu, Y., Yu, Q.\ 2012, ApJ, 761, 5
 (arXiv:1210.4019) 
\end{references}
\end{document}